# MobChain: Three-Way Collusion Resistance in Witness-Oriented Location Proof Systems Using Distributed Consensus

Faheem Zafar[1], Abid Khan[2, *], Saif Ur Rehman Malik[3], Adeel Anjum[1], Mansoor Ahmed[1]

[1]Department of Computer Science, COMSATS University Islamabad (CUI), Islamabad, Pakistan

[2]Department of Computer Science, Aberystwyth University, Aberystwyth, SY23 3DB United Kingdom

[3]Cybernetica AS, Talliann, Estonia

faheemiiui@gmail.com, abk15@aber.ac.uk, saif.rehmanmalik@cyber.ee,

adeel.anjum@comsats.edu.pk, mansoor@comsats.edu.pk,

## Abstract

Smart devices have accentuated the importance of geolocation information. Geolocation identification using smart devices has paved the path for incentive-based location-based services (LBS). A location proof is a digital certificate of the geographical location of a user, which can be used to access various LBS. However, a user's full control over a device allows the tampering of location proof. Initially, to resist false proofs, two-party trusted centralized location proof systems (LPS) were introduced to aid the users in generating secure location proofs mutually. However, there were three problems with centralized architecture: *Firstly*, LPS must depend on a centralized trusted authority. *Secondly*, the proposed centralized architectures were too complex to work. *Thirdly*, the two-party protocols suffered from the collusion attacks by the participants of the protocol. Consequently, many witness-oriented LPS have emerged to mitigate collusion attacks in two-party protocols. However, witness-oriented LPS presented the possibility of three-way collusion attacks (involving the user, location authority, and the witness). The three-way collusion attacks are inevitable in all existing witness-oriented schemes. To mitigate the inability to resist three-way collusion of existing schemes, in this paper, we introduce a decentralized consensus protocol called as *"MobChain"*, where the selection of a witness and location authority is achieved through a distributed consensus of nodes in an underlying P2P network of a private blockchain. The persistent provenance data over the blockchain provides strong security guarantees, as a result, the forging and manipulation become impractical. MobChain provides secure location provenance architecture, relying on decentralized decision making for the selection of participants of the protocol to resist three-way collusion problem. Our prototype implementation and comparison with the state-of-the-art solutions show that MobChain is computationally efficient, highly available while improving the security of LPS.

## 1. Introduction

Location-based services (LBS) have reformed the business by offering services based on the current geographical location of users. The importance of smart devices to provide a person's accurate geographical location is increased for the consumption of location-based services. Consumer-based applications like locator services (nearest restaurants, stores, and ATM, etc.), location-based contents (games, news, and weather updates, etc.), location-based social networks (LBSN), vehicle route guidance using Google-maps have been possible through smart devices [1]. Other than consumer-based services in businesses like courier services; pharmaceutical distributors, etc. location information can improve business operations in the field using smart devices. The real challenge is to prove that the person was physically present at the reported location and time. Location data provided by the individuals themselves cannot be trusted. The user may cheat, to earn the incentives or to hide noncompliance to job responsibilities. With the advent of the location proof system (LPS), it is now possible to achieve trustworthy proof of user physical presence on a location using smart devices. LPS aids in the generation of location-proof (LP) [2], which is a digitally signed asset certifying the user presence on geographical location at a specific time instance. LP contains the user identity, location coordinates, and timestamp, which is verifiable in a secure manner. To better understand the importance of the LPS, the following are a few real-world examples of its application: *(a) A utility store might offer special discounts on customer's loyalty. To claim the discounts, customers may use their smart devices to get proof of every visit from the store and present them in the future as the history of the visit. (b) A special task force assigned to a critical mission in the enemy area may be asked to keep a copy of their location traces such that the commanding center may perform post-mission analysis of navigation. (c) Transportation incentives for following the environment-friendly routes. (d) Courier Service delivering mobile bills may ask the employees to get location proofs to ensure that they have delivered the bills to the customer's location ensuring the quality of the operation. (e) Pharmaceutical companies can utilize the location proofs by medical raps and sales-person to ensure that they are visiting hospitals and doctors* [3]. *(f) Construction companies can benefit by utilizing location proofs to ensure that the engineers are visiting the sites. (g) Location-based access control* [4].

LPS uses the *"localization"* concept by which a smart device can report its location relative to the co-located device or cellular tower or using a global positioning system (GPS) etc. [5]. For localization, proof generation systems are designed to utilize Bluetooth [6], GPS [7], Wi-Fi infrastructure [8], and infrared [9]. Other techniques like distance bounding protocols [10] and proximity [5] [11], cellular tower triangulation, mobiles triangulation, IP address tracking [3], and audio-based positioning [9], etc. were used to generate location proofs in the past. The initial system's design was based on a centralized architecture with a trusted third party to validate the user's location claims. However, centralized LPS were prone to a single point of failure and suffered performance issues in peak load times. Limitations of centralized LPS were mitigated by introducing a distributed LPS architecture [8]. Distributed LPS started with a Two-party protocol (involving Prover, Location Authority) and have gradually evolved to witness-oriented 3-Party (Prover, Location Authority, Witness) and Multi-Party (Prover, Location Authority, Multiple Witnesses) protocols for a location-proof generation. However, distributed LPS are prone to collusion attacks. Many approaches have been used to mitigate collusion attacks using outlier detection techniques [6], entropy-based trust model along with localization using distance bound protocol [8] and witness asserted location proofs [12]. Later, the LPS have evolved to support location provenance. Location provenance chaining tracks the location history of the traveled path in a secure manner by keeping the chronological order of locations visited. The chronological order of location information is of great importance to increase the reliability of location proofs. Location provenance chaining can be also used to detect attacks like backdating or future dating. Furthermore, a trust evaluation system can also be designed over a location provenance chain. For example, a trust evaluation mechanism can be based on spatio-temporal correlation, where time difference and distance between two consecutive location proofs can aid in the determination of false proof [13]. Consider a case, where a user can successfully get location proof $LP_1$ for location $A$ at time instance $T_1$ and $LP_2$ for location $B$ at time instance $T_2$. Now $(T_2 - T_1)$ is short enough that practically the user cannot travel from $A$ to $B$ in this much short time.

Nowadays, blockchain-based provenance schemes from a resilience and security perspective are increasing in popularity [14]. Blockchains [15] [16] is well known as immutable distributed ledgers. Immutability and implicit record chaining in blockchain have enabled its application in systems like recording the chain of custody in supply-chain management system [17] [18], data provenance management in cloud storage [19], privacy-preserving smart contracts [20] [21], decentralized file storage [22] [23], etc. However, the application of blockchain for location provenance management to record the chronological order of the travel history securely is getting the attention of the research community. To the best of our knowledge, Giacomo *et al.* [24] proposed the first decentralized LPS using blockchain for P2P overlay schemes. They used blockchain for storing location provenance data to resist against backdating and future dating attacks. Later, decentralized LPS such as [25] and [26] were also proposed based on the P2P network of blockchain.

The trustworthiness of location proof demands a cheat-proof system [27]. Because the user has full control over his smart device, therefore he can lie about its position or can tamper the proof after its generation. Another possibility is that the participants may collude to generate fake proof of physical presence. In the existing literature outlier detection techniques, entropy-based trust model, and witness asserted location proofs have been used to mitigate collusion attacks to some extent. However, to the best of our knowledge, the three-way collusion is not addressed by any existing witness-oriented schemes. Three-way collusion occurs when a *prover*, *witness*, and *location authority (LA)*, all three are malicious and may collude to generate a false proof of location. After the literature review, two types of weaknesses are identified such that if any of these exists then three-way collusion is inevitable:

1. **Participant selection decision-control lies with participants of the protocol:** Participant's selection control for proof generation lies with the user who will always choose the colluding participant to cheat. Xinlei *et al.* [8] proposed the STAMP scheme, which relies on witness selection on the peer discovery mechanism provided by the underlying communication technology of the user's device. Similarly, Giacomo *et al.* [24] scheme allow the user to select the peer for assistance in a proof generation.
2. **Participant selection decision control lies with the single party:** User has direct communication with LA, and both can collude to generate false proof. Even in the witness-oriented scheme, witness selection control either lies with prover or LA. So, if both the prover and LA are colluding, it will be easy for them to choose a colluding witness. In Rasib *et al.* [3] scheme, LA is the stationary entity on a site, which randomly chooses a witness from the "witness list" for aiding the user in a proof generation. Furthermore, the puppet witness attack and wormhole attacks are easy in this scenario.

Our proposed MobChain scheme is designed to mitigate these weaknesses and to provide better resistance against a three-way collusion problem in existing schemes.

The main contributions of this paper can be summarized as follows:

1. To the best of our knowledge, MobChain is the first three-way collusion resistant witness oriented LPS. The design of MobChain is inspired by the blockchain, where the P2P network is responsible for decentralized decision making for the selection of participants of a location proof protocol. Furthermore, the underlying P2P network is also responsible for maintaining location provenance in a decentralized fashion.
2. We have developed a *proof-of-concept (POC)* application using the AKKA [28] toolkit to compare its performance with the WORAL framework [12]. Experimental results show that the performance of the proposed scheme even with the overhead of decentralized decision making, still competes with WORAL.
3. MobChain enables the deployment of multiple location authorities to remove a single point of failure, which was ineffective in existing schemes because the control of location authority selection was with the user. Secondly, MobChain mitigates the denial of service (DoS) attacks. For example, consider a scenario in which a single location authority is flooded with proof requests then load balancing mechanism of MobChain can distribute the requests to other available location authorities on the site.

The rest of the paper is organized as follows: Section 2 provides a detailed overview of the related work in this domain. In Section 3 we provide an overview of the proposed MobChain scheme. In Section 4, we highlight the threat model and the assumptions about the adversary. The proposed

MobChain architecture is presented in Section 5. In Section 6 we provide the detailed security analysis of the proposed scheme. Our experimental evaluation of the proposed architecture is provided in Section 7. In Section 8 we conclude this paper and provide directions for future research in this domain.

## 2. Related Work

In this section, we review the existing work on LPS. Generally speaking in location proof systems, a *"prover"* is the main entity that wants to generate secure location proof for his physical presence at a particular time instance through secure localization [2] [8]. Localization is a mechanism by which a smart device can tell "Where am I?" relative to some other smart device, map, or global coordinate system [5]. Location-aware schemes rely on either software-based or hardware-based localization techniques. Gabber *et al.* [29] have incorporated multiple channels (GPS, cellular telephony, Caller-Id, satellite, etc.) to monitor the movement and location of smart devices. However, it has been proved later that a malicious entity can bypass such a multi-channel combination approach [30]. For example, GPS signatures [31] were prone to spoofing attacks [32]. Bauer *et al.* [33] have discussed the vulnerabilities of wireless-based localization approaches against non-cryptographic attacks using a low-cost antenna. Gruteser *et al.* [34] have proposed an anonymity-based privacy-preserving localization technique. This scheme is based on middleware to adjust the location information along spatial-temporal dimensions for a centralized location broker service. Zugenmaier *et al.* [35] have introduced the "*location stamps*" concept utilizing cell phones. Dominik *et al.* [36] have proposed a secure and tamper-resistance location proof system based on visual features and image recognition without overburdening the user.

Researchers are also exploring hardware-oriented localization schemes. Hardware-based localization techniques include measuring signal attenuation [32], measurement of round-trip time (RTT) [10], voice signatures [37]. However, these approaches have failed to provide secure localization under adversarial settings. In [32], the secure positioning of wireless devices under adversarial settings has been discussed. Analysis of positioning algorithms (including received signal strength (RSS)), ultrasound time-of-flight (TOF), radio TOF, civilian GPS) against position and distance spoofing attacks is performed providing the vulnerabilities details. Signal attenuation techniques suffer from channel noise and the constraint of line-of-sight makes them difficult in practical scenarios [12]. Saroiu *et al.* [38] have used a trusted hardware trusted platform module (TPM) and virtual machine-based attestation to make the sensor readings trustable. Similarly, Gilbert *et al.* [39] have proposed a TPM based trustworthy mobile sensing platform to provide data integrity and privacy protection.

Furthermore, Luo *et al.* [30] have proposed six design goals for a proactive location proof system providing privacy protection. Saroiu *et al.* [40] have devised a Wi-Fi-based protocol, where Wi-Fi access points (AP) aids the prover in the generation of trusted location proofs. However, their scheme is prone to collusion attacks as AP and prover can collude to generate fake location proofs. User privacy has been the primary concern in [40] as the real identity of the user was exposed to AP. Authors in [40] have described the security properties of secure location proofs and have discussed the applications where LBS with incentives provide a motive for users to lie about the location. The VeriPlace [41] is a privacy-preserving LPS with collusion resistance support. However, the assumption of the short interval between location proofs for collusion detection makes the *VeriPlace* vulnerable. If the interval between two chronologically close location proofs is not close enough, then the *VeriPlace* will treat them as suspicious. Therefore, *VeriPlace* puts the burden on the user to have frequent location proofs. All these schemes have ignored the chronological order of location proofs. Hassan *et al.* [42] have designed a scheme for location proofs with wireless access points as location authorities and co-located smart devices designated as witnesses for proofs endorsement through Bluetooth. This scheme has removed the dependency on the trusted third parties. Nonetheless, location provenance was maintained by [42] to record the chronological order of location proofs, which was missing in previous schemes. All the above schemes are under the category of centralized architectures.

In the distributed model, Davis *et al.* [43] have devised a scheme to generate location proofs with the help of smart devices within proximity. However, the scheme [43] was not collusion resistant. Zhu *et al.* have proposed an LPS called *"APPLAUS*" [6] (i.e A Privacy-Preserving LocAtion proof Updating System), which has utilized the bluetooth technology allowing the co-located devices to generate the location proofs mutually. *APPLAUS* has utilized *"pseudonyms"* for privacy preservation. However, communication overhead is introduced due to the generation of dummy proofs periodically. Moreover, to improve the security of *APPLAUS*, the authors have devised a collusion detection mechanism based on the ranking and correlation clustering approaches. However, the adopted collusion detection mechanism was later proved to be energy inefficient and a successful detection ratio > 0.9 was possible only when collusion percentage is < 0.1. Ananthanarayanan *et al.* [44] have introduced a framework called "StarTrack" enabling tracks of information holding data about a person's location, time and metadata. However, this concept was quite like the location provenance chain as data recorded in the time-ordered sequence. Besides, no security measures were taken and thus leaving the scheme vulnerable to malicious user manipulations. Gonzalez-Tablas *et al.* [45] have presented the notion of "*Path-stamps"*, extending the concept of *"location-stamps"* [35] by recording the history of the visited location's proofs in a hash chain. Rasib *et al.* [3] have relied on a WiFi-enabled smart device to generate the location proof and have also proposed the formal requirements for the design of LPS. The scheme [3] treated the location authority to be malicious and empowered the LPS with witness endorsement. Additionally, the authors have highlighted the possible attacks and devised a trust score mechanism to evaluate the reliability of witnesses providing the collision resistance. However, none of these schemes described the requirements for a secure location provenance mechanism formally. In *OTIT* [46], for the first time, the requirements of the secure location provenance have been formally defined. Furthermore, the authors have also performed a comparative analysis of different techniques used to maintain a provenance chain. The comparison has been based on provenance generation time, sequential verification time, sparse verification time, and space requirement. Wang *et al.* [8] have proposed *"STAMP"* which is a spatial-temporal provenance assurance with the mutual proofs scheme. STAMP ensures a user's privacy while providing the integrity and non-transferability of location proofs. To guard against collusion attacks, an entropy-based trust model is utilized. Furthermore, STAMP reduced the dependency on multiple trusted parties to a single semi-trusted party i.e. certification authority (CA). The scheme has also supported the granularity level control for the exposer of location information to the verifier by the user. STAMP is the first scheme to deal with two collusion attacks i) User-A physically present at a target location generates a false proof for a User-B by masquerading ii) *Terrorist Fraud Attack*: two malicious users colluding to generate fake proof of location for each other. To prevent terrorist fraud attacks, bassard-bagga distance bounding protocol has been used with a trade-off on performance in STAMP. Later on, Hasan *et al.* [12] proposed *"*Witness ORiented Asserted Location provenance *(WORAL)",* which is a distributed witness-oriented secure location provenance framework for mobile devices. WORAL is a complete working system build by integrating the *asserted location proof (ALP)* (proposed in the scheme [3]) and OTIT [46] model for managing secure provenance. WORAL has provided collusion models and corresponding threats. Furthermore, the authors have also claimed that the system is only 12.5% vulnerable because of the inability to resist three-way collusion. WORAL has evaluated the protocol based on characteristics including the proof generation time, maximum distance threshold (depending on localization technique), proof size, number of participants of protocol, collusion detection rate. WORAL has established the vulnerability matrix to ensure that fake proof generation is not possible in any scenario. For privacy-preservation, *crypto-ids* has been used by WORAL, such that the many-to-one relationship holds between crypto-ids and the real identity of the user. Giacomo *et al.* [24] have proposed the first decentralized location proof system by using blockchain for P2P overlay schemes. However, participant selection control is in hands of the prover, therefore, decentralization standalone still makes their scheme vulnerable to collusion attack. Proof generation protocol of the scheme [24] allows the direct communication between the prover and responding peer, therefore, the possibility exists that both can collude. However, the scheme [23] can resist backdating and future dating attacks. Amoretti *et al.* [25] have proposed a blockchain-based LPS considering both static and dynamic entities. Nasrulin *et al.* [26] have also proposed the decentralized location proof system and evaluated its performance and security by developing the POC for supply-chain management. Moreover, Wenbo *et al.* [47] have discussed the cheating possibilities of users against the location verification mechanism used by FourSquare. Authors crawled the website and analysis were performed on the crawled data to highlight the vulnerabilities, exploitable for cheating on the user's location. Another dimension of research based on location information is quantum-based geo-encryption [48], which enhances the security of a traditional cryptosystem by introducing a "geolock" in which the encrypted data can only be decrypted at a specific location. The motive of quantum-based geo-encryption is to reduce the chances of spoofing attacks to zero as the adversary gets away from the targeted location. Brassil *et al.* [49] have proposed the robust location authentication mechanism while relying on the femtocells of 802.11x access points. Basic idea was to analyze the traffic signatures for location verification to make the scheme device-independent and carrier independent (3G, LTE, *etc.*), which was efficient due to the use of the non-cryptographic technique for location verification. Idrees *et al.* [51, 52] have proposed secure provenance schemes in a distributed environment using aggregated signatures. Furthermore, the authors did not assume any transitive trust among consecutive colluding users, which is a stronger security model compared to previous studies. However, the proposed scheme is not intended for providing secure location proofs. A detailed review of trustworthy data using provenance in various domains is provided by the authors in [53]. The authors have explored the notion of trust using secure provenance in various domains such as wireless sensor networks, cloud computing, and databases. A comparative analysis of essential security properties is also provided for various secure provenance schemes in these domains. A similar study recently has identified the requirements and challenges for location proof systems in []

## 3. Threat Model

In our threat model, we discussed the adversary's roles and capabilities. We also discuss the possible attacks by individuals and colluding malicious parties of the system. The primary assets of the MobChain that are vulnerable to attacks by malicious participants include:

- **Decision Block:** Block generated by a consensus of supervisor nodes for the selection of location authority and witness to aid the prover for a proof generation.
- **Decision Blockchain:** Blockchain recording the decision blocks generated by consensus of supervisor nodes in chronological order.
- **Location Proof:** A proof of the presence of the user at a location with an exact timestamp.
- **Location Provenance Blockchain:** Blockchain recording all the past location proofs in a chain to keep their chronological order intact.

To the best of our knowledge, currently, no location proof system has considered a three-way collusion, i.e. all three participants (prover, witness, location authority) involved in a proof generation can be malicious and colluding at the same instance. However, for the security of MobChain, we have considered the following possible attacks discussed in the literature [3] [8] [12]:

1. **False Presence:** A malicious prover may want to get a fake proof of location without being physically present on the claimed location.
2. **False time (back-dating, future-dating):** Prover tries to get the proof for his true location with timestamp different from the time of visit. Participants of the protocol collude to generate the location proof for the prover with a past timestamp in the back-dating attack and with a future timestamp in the future-dating attack.

3. **Sequence Alteration:** In a sequence alteration attack, the prover tries to present a false travel path by changing the chronological order of location proofs.
4. **Implication:** Participants of the protocol dishonestly prove the physical presence of the prover at the location to victimize the prover.
5. **False Endorsement:** In the witness-oriented model, the witness colludes with the prover to falsely endorse that prover is physically present on the claimed location.
6. **Presence Repudiation:** User tries to deny his presence on a location at a time instance for which location proof has been generated.
7. **Proof Tampering:** A legitimate old proof's timestamp or location information may be modified to present it as new proof.
8. **Puppet Witness Attack:** Location authority and prover may collude and create a puppet witness to falsely endorse the proof of location or relay the request to a remote witness not co-located to prover at the time of visit.
9. **Wormhole Attack:** Any entity of the system physically present on the desired location may impersonate the prover and generate the location proof for him. It is not necessary that the prover has shared his secret keys with the impersonating party but might have established a covert communication channel and the impersonator might be relaying the messages of proof generation protocol to the prover. Wormhole attack is also known as a *terrorist fraud attack*.

**3.1 Assumptions:** We make the following assumptions about the adversary:

1. Participants of the system are well known to the system and they do not share their private keys.
2. Users do not have multiple identities to launch a Sybil attack.
3. Smart devices are not shared with other participants.
4. At least one witness is available on the location of the visit.
5. No entity in the system will be able to compromise 51% of the supervisor nodes [25].

**3.2 Goals**: MobChain has the following goals:

- Provide three-way collusion resistance while ensuring protection against known attacks over existing schemes.
- Measure the impact of introducing distributed consensus in location proof generation protocol.
- Analyze the storage requirements for peers (supervisor nodes) for maintaining the blockchain.

## 4. MobChain

**Overview and motivation:** Building a secure location proof system capable of collusion resistance is a real challenge. Since the participants of the protocol are not trustable and are in full control of their smart devices. In the given circumstances, if participant selection control lies with either the prover or the designated authority, then collusion is inevitable. To mitigate the "three-way collusion" problem, we propose a decentralized secure location provenance architecture called "MobChain", which extends the design of WORAL [12], by introducing decentralized decision making for the selection of participants (LA, Witness) for the location proof generation. WORAL is a witness-oriented location proof scheme capable of providing the location provenance. WORAL treated the location authority as malicious along with witness and prover.

Since in practical situations, all three participants (prover, location authority, and witness) can be malicious and colluding with the prover, therefore, a location proof system cannot guarantee the reliability of location proofs assuming the location authority as trusted. Therefore, WORAL is assumed to provide better security in the context of location proofs. However, the weakness of the WORAL scheme is the strong assumption that all participants will not be colluding at the same time in a location proof protocol which is termed as "three-way collusion" in literature. The inability to resist three-way collusion by WORAL serves as a base for our problem statement. In the presence of a three-way collusion, fake proof of location can easily be generated.

To improve the security of location proof systems, our contribution is the proposal of two design principals to mitigate these weaknesses:

i) Separation of participant selection control from location proof generation protocol
ii) Decentralization of control decisions (selection of participants, the addition of location proofs in provenance chain, validation of location proofs on the request of the third party).

By separating the participant selection control from the location proof generation protocol makes the collusion hard for malicious participants because a third party will decide that who will be assisting the prover for a proof generation. However, delegating the participant selection control to a single third-party still has a high probability to allow collusion if assumed malicious. Therefore, we adopted the decentralized control-decision strategy in our proposed scheme to make the collusion harder for participants. Decentralized decisions will demand the malicious participants to compromise the majority of the decision-makers to get the decision in their favor to generate fake proof that is practically hard [25].

In MobChain, the provers and witnesses are mobile entities visiting a location temporarily, while the location authority is a static entity designated on the location permanently. Therefore, the location of location authority is pre-known while mobile entities visiting the location report their location to MobChain P2P network (i.e., the admin layer established by supervisor nodes). In MobChain, witness, prover, and location authority all are the participants of the location proof protocol and are part of the service layer. Any mobile entity visiting the location can be in the role of prover and witness. Once a mobile entity requests for the proof of location, it becomes the prover for that request while all other entities co-located on the location are eligible to become a witness for that prover. For some other instance of time, this witness can become the prover on requesting the proof of its location and the prover can become the witness for him in a location proof protocol. Since the location authority is the static entity designated by the system, therefore its location is pre-known and is involved in the location-proof generation to ensure that prover and witnesses are physically present on the location.

Furthermore, all the supervisor nodes in the admin layer are eligible for receiving the location proof requests from visiting mobile entities. The Admin layer node on receiving the location request from the visiting entity broadcasts the message in the admin layer to let other nodes know about the presence of a mobile entity. Mobile entities keep the admin layer updated about their presence and exact location through a periodic ping mechanism to remain eligible for the selection as a witness. Because in witness selection criteria, two parameters are considered for mobile entities i) uptime and ii) number of requests entertained. *'Uptime'* for mobile entities is calculated using the time difference of first location request and last ping time. If the mobile entity's ping request is not received within a certain time, then it is removed from the list of available mobile entities. Later, when the mobile entity comes back on the location, then it is considered as a new entity.

Another promising aspect of MobChain is utilizing the blockchain capabilities to support location provenance. Location provenance provides the history of locations visited by maintaining the tamper-evident chain of all location proofs generated in the past while keeping their chronological order intact. Blockchain is a public distributed ledger, which stores the transaction data in a modification-resistant chain [14]. Since blockchain is implicitly tamper-resistant and maintains the chronological order of data inserted therefore it is an ideal candidate to maintain location provenance. Additionally, blockchain removes the single point of failure as all peers of the blockchain hold the data and can validate it on request. Peers of the P2P network of the blockchain establishes distributed consensus before committing any data over the blockchain. In non-blockchain based location proof systems, location provenance is either maintained by the location authority or saved on the prover's smart device. If location provenance is maintained on the user's smart device only, then in case of damage or stolen, all location proofs will be lost. If location provenance is maintained by location authority, even then it remains the single point of failure and can become a bottleneck in case of high loads for location proofs verification. To overcome these limitations, blockchain is ideal for maintaining location provenance. However, the primary difference between blockchain and MobChain is that in traditional blockchains, the decentralization focus is on establishing a distributed consensus about whether to make any new data block part of the blockchain or not. While in MobChain, the decentralization primary focus is on establishing distributed consensus for control decisions including:

- Approval of witness and location authority for the prover to start location proof generation.
- Making approval decision block, part of the decision blockchain (maintained by the admin layer of MobChain).
- Making generated location-proof, part of a location-provenance chain (maintained by the admin layer of MobChain).
- Validation of old location proofs on the request of the third-party involving a decision blockchain and location-provenance chain.

For location-privacy preservation, we adopted the crypto-id based pseudonyms approach of WORAL for MobChain. Supervisor nodes can only validate the prover based on its "crypto-id" to ensure that it is a valid user of the system. However, supervisor nodes cannot link their location information to the personal identity behind that "crypto-id" to violate the privacy of the user.

## 4.1 Architecture Overview

Before the explanation of the MobChain architecture, we need to elaborate on certain abbreviations and terminologies. *The prover* is the entity of the system requesting the location proof of his physical presence at a specific location. *Location authority (LA)* is the stationary entity at a location that aids the prover in location proof generation. *The witness* is a co-located entity who asserts the location proof generated by the prover and location authority to approve that prover is physically present on the claimed location. *Worker node (WN)* term points to a general category that encompasses both the location authority and witnesses. The *Supervisor node (SN)* refers to the individual peer which is part of a P2P network of the MobChain admin layer. The *Request receiving supervisor node (RRSN)* is the supervisor node who receives the *proof request (PReq)* from the prover and initiates the distributed consensus for location proof protocol participant's selection. However, all the supervisor nodes in the admin layer can become RRSN for different provers. *RRSN* is not a designated role for any specific supervisor node. We labeled request receiving supervisor node *RRSN* to differentiate it from other supervisor nodes in the schematic description of MobChain working in section 5.2. *Decision block (DB)* is the final decision message created by *RRSN* when distributed consensus protocol ends. *RRSN* then sends the approval message which includes the decision block reference such that it allows the prover, location authority, and witness to initiate the location proof generation. Table 1. summarizes the abbreviations

Table 1. Abbreviations and their description

| Abbreviation | Term | Description |
|---|---|---|
| *LA* | Location Authority | The designated stationary entity on each site who aids the prover in location proof generation while ensuring that witness and prover are physically present on the location. |
| *WN* | Worker Node | Worker node refers to LA(s) and witnesses who provide services to prover for the generation of asserted location |
| | | proof. |
| *SN* | Supervisor Node | Admin layer peers are called supervisor nodes. These nodes are responsible for decentralized decisions. |
| *RRSN* | Request Receiving Supervisor Node | Supervisor node (*SN*) who initiates the distributed consensus protocol on receiving the proof request from prover. Any of the supervisor nodes can receive the location proof request. *RRSN* is a label to differentiate the role of request receiving supervisor node from other supervisor nodes in the system. For example, Prover1 requests SN1 for location proof then SN1 is considered *RRSN* while at that same time Prover2 request SN2 for location proof then SN2 is considered RRSN in for the Prover2 request. |
| *CWN* | Chosen Worker Node | *CWN* refers to *LA* and Witness chosen by the admin layer through distributed consensus against the proof request of the prover. |
| *PReq* | Proof Request | Proof request message symbol. |
| *DAM* | Decision Acknowledgement Message | During consensus protocol execution, every *SN* decides the witnesses and *LA* for prover and informs the *RRSN* about his choice by sending a decision acknowledgment message. |
| *DB* | Decision Block | The Decision block is the final message generated by *RRSN* on the completion of distributed consensus. This decision block is made part of the decision blockchain. |
| *AMsg* | Approval Message | On completion of distribution consensus, the prover is provided with the approval message containing the decision block id, chosen witness, and location authority. |
| *LPReq* | Location Proof Request | Prover requests the location authority to aid him in proof generation using an approval message as an authentication token. |
| *LP* | Location Proof | *LP* is a digital certificate generated by *LA* approving the physical presence of prover at the location on a specific time instance. |
| $AReq_{LP}$ | Assertion Request (of Location Proof) | The location authority requests the witness for *LP* assertion by sending *AReqLP*. |
| *AR* | Assertion Response | Witness after successful verification of the prover's location and validation of *AReqLP* sends back assertion response (*AR*) to location authority. |
| *ALP* | Asserted Location Proof | Witness asserted *LP* is termed as *ALP*. |
| *VReq* | Verification Request | To ensure the validity of *ALP*, prover issues a verification request *VReq* to witness. |
| *VR* | Verification Response | Witness responds the prover by *Yes/No* after validation of *VReq* by sending the message *"VR"* |
| $ACK_{ALP}$ | Acknowledgment Message | On successful verification of asserted location proof, the prover sends the acknowledgment $ACK_{ALP}$ to location authority to end the protocol. |

In section 1.1, we identified the two primary weaknesses of the existing schemes which result in an inability to resist three-way collusion. Therefore, to enable resistance against three-way collusion we laid the foundation of MobChain on two design principals:

i) Separation of participant selection control from location proof generation protocol
ii) Decentralization of control decisions

To incorporate these design principals, the underlying P2P network of MobChain is virtually organized into two layers:

1. **Admin Layer:** All kinds of control decisions are the responsibility of the admin layer. Commodity devices in the admin layer form a virtual cluster of supervisor nodes. Supervisor nodes take proof requests from prover and all control decisions are taken through a distributed consensus mechanism within the admin layer. Primary operations of the admin layer are:
   - **a.** Distributed Consensus (decentralized decision for the selection of worker nodes co-located to prover for assistance in location proof generation).
   - **b.** Final validation of location proof generated and adding it to blockchain to build location provenance.
   - **c.** Auditor role to verify the location proof requested by the third party.
2. **Service Layer:** In MobChain, witness, prover, and location authority all are the participants of the location proof protocol and are part of the service layer. Location authority is the static entity permanently designated at the location while witnesses and provers are mobile entities, who may visit the location temporarily. Any mobile entity visiting the location can be in the role of a prover and witness. Once a mobile entity requests for the proof of location, it becomes the prover for that request while all other entities co-located on the location are eligible to become a witness for that prover. At some other instance of time, this witness can become the prover on requesting the proof of its location and the prover can become the witness for him in location proof protocol. Since location authority is the static entity designated by the system, therefore its location is known in advance. Furthermore, it is responsible for ensuring that the prover and witnesses are physically present at the location. All mobile entities visiting the location connect to supervisor nodes in the admin layer to let the system know their presence at the location. No permanent connection is required with the admin layer by these entities. However, a periodic ping mechanism keeps the admin layer informed about their presence and location. Once any of these mobile entities request the system to provide a location proof, it becomes the prover and other co-located entities are considered available witnesses. RRSN pings the witness selected after distributed consensus to ensure that is available and then sends the approval message to prover to initiate the location proof protocol with the chosen witness and location authority.

Two types of blockchains are maintained by the admin layer of MobChain:

i) **Decisions Blockchain**: Tracks all decisions taken by the admin layer keeping their chronological order intact with timestamps. Blocks of the decision chain will serve for validation of location proofs and cross-checking of the witness elected by the admin layer and the actual witness who aided in a proof generation. It will also provide protection against future and backdating attacks and three-way collusion for a fake proof generation or representation of valid old proof after tampering.

ii) **Location Provenance Blockchain**: All location proofs of the prover will be recorded in chronological order in the location provenance chain.

All supervisor nodes will be deployed at multiple geographical locations and the initial P2P network will be established. Later, public addresses of supervisor nodes will be published, so that worker nodes can join the network. Worker nodes will authenticate with the supervisor node to join the P2P network of MobChain. Geolocation of the newly joining worker node will be broadcasted among supervisor nodes thus a common list of available worker nodes with their geolocations is established in the admin layer of MobChain. Geolocations of worker nodes can be refreshed on-demand and at regular intervals. Supervisor nodes will use geolocations from worker nodes list to find prover's co-located worker nodes to establish a distributed consensus.

## 4.2 Distributed Consensus

Decentralizing the decisions demands a mechanism to establish a consensus on the same value by the underlying P2P network. In MobChain, the decentralized decision of selecting worker nodes (*LA* and Witness) for the prover is achieved through a specialized consensus mechanism as described in Figure 1. The consensus mechanism is started by *RRSN* on receiving the proof generation request. *RRSN* generates the specialized message including the proof request from prover and broadcasts it in the admin layer. On receiving the specialized request message from *RRSN*, all supervisor nodes evaluate prover's co-located workers (witnesses and location authorities) using geo-coordinates of prover and available worker nodes. Once co-located workers (LA and Witness) are evaluated, a signed decision message (*CWN*) is sent to *RRSN* by every supervisor node. *RRSN* will wait until the consensus threshold satisfied i.e. $(N/2) + K$ signed decision messages agreeing on the same witness and location authority are received. $N$ is the number of supervisor nodes and $K$ is the at least percentage of decision messages above 50% agreeing on the same witness and location authority. The consensus threshold value is a tradeoff between the reliability of the decision and the decision time. The higher the consensus threshold, the more reliable will be the decision but the higher will be the decision time. Therefore, we have chosen $(N/2) + 1$ consensus threshold value to balance this tradeoff. Once the consensus threshold is accomplished, the *RRSN* will generate a decision block by including all the signed messages received. The finalized decision block will be then broadcasted in the admin layer by *RRSN*. The decision block generated is passed to the prover and will be used in generated location proof. All supervisor nodes will validate their signed message and the original request from prover in the decision block to make it part of the decision blockchain. If validation of the decision block fails, then the supervisor nodes will discard it. If the decision block is discarded by the major of the supervisor nodes, then location proof generated using the decision block id will be treated as fake and thus discarded by the admin layer. Only those location proofs will be made part of the location provenance chain for which valid decision block will be present in decision blockchain.

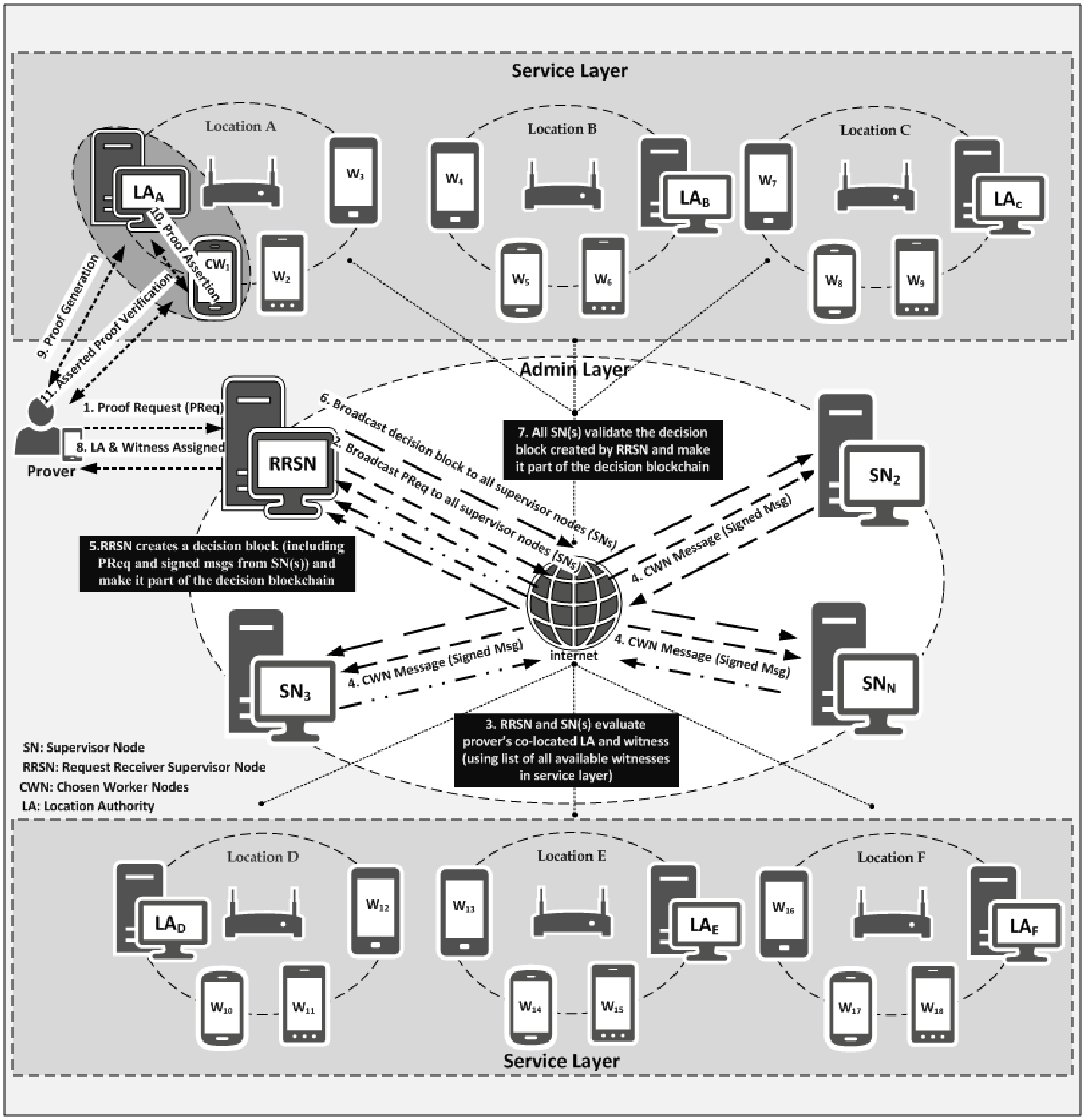

Fig 1. Proposed MobChain Architecture

### 4.3 Schematic Description of Distributed Consensus and Location Proof Generation in MobChain

Working of MobChain is divided into 2 phases

i) **Decentralized Decision on Witness and LA Selection (Distributed Consensus) Phase:** On prover's request, *RRSN* initiates the protocol over the admin layer such that witness and *LA* are elected through distributed consensus of supervisor nodes.

ii) **Witness-Oriented Location Proof Generation Phase:** Once *LA* and witness are intimated to assist the prover; the actual location proof generation phase starts.

To explain the working of location proof generation protocol, we are using the schematic description of the messages communicated in each step of both phases.

#### ***4.3.1 Decentralized Decision on Witness and LA Selection (Distributed Consensus) Phase:***

Protocol initiates with *PReq* sent by prover to any supervisor node which is represented as *RRSN*:

$$Req = [ID_P, T_P, L_P] \quad (1a)$$

$$PReq = [Req, Sign_P(Req)] \quad (1b)$$

In expression (1a), $ID_P$ is the unique identifier of the Prover, $T_P$ is the timestamp prover's smart device, $L_P$ is the current location of the prover, and $Sign_P(Req)$ will protect against repudiation of the prover. To present that prover at location $L_P$ has requested for proof generation, *RRSN* constructs a message *PReq'* and broadcasts it to all other *SN*(s):

$$PReq' = [PReq, T_{RRSN}, Sign_{RRSN}(T_{RRSN}), ID_{RRSN}] \quad (2)$$

In expression (2), $ID_{RRSN}$ is the unique identifier of *RRSN*, $T_{RRSN}$ is the timestamp of the *RRSN* machine when it received the *PReq*. $T_{RRSN}$ indicates the freshness of the message and $Sign\ (T_{RRSN})$ helps the other *SN*(s) to ensure that the request is from the known valid supervisor node of the system. *PReq'* thus will help to mitigate replay attacks as any message with $T_{RRSN}$ older than a certain threshold duration will be discarded. On receiving the message from *RRSN*, every *SN* validates the *PReq'* and then find the prover's co-located witness and *LA*. For this purpose, SN(s) computes the priority for all available witnesses and *LA*(s) registered in the network using the below expression,

$$Priority = \frac{\text{No. of requests entertained by [Witness/LA] x Uptime}}{\text{The distance between Prover and [Witness/LA] } (\leq \text{Short Range Communication Technology})} \quad (A)$$

In expression (A), the number of requests entertained is assumed to be greater than 0 and will help in rotating the chance of participation by all entities in proof generation whereas uptime is the measure of the reliability of the witness and location authority. To elaborate on the impact of priority calculation expression on the security of MobChain, suppose the *RRSN* supervisor node is compromised and colludes with prover to make the centralized decision in favor of a malicious witness who is willing to collude with the prover. If we see the priority calculation equation, the number of requests entertained (by witness & location authority) and uptime are two parameters which enable honest supervisor nodes to prevent the MobChain against compromised supervisor node. Since the number of requests entertained by the witness and the location authority is the count of old location proofs in which witness, and location authority individually participated. Therefore, several requests for entertained information cannot be modified by compromised supervisor node in the admin layer. The second parameter is the uptime of witness and the location authority which is the difference of the first request to the admin layer (to notify its presence and location) and the last ping time. Therefore, the uptime value of any witness and the location authority cannot be manipulated by any single supervisor node in the admin layer. Another possibility for a compromised supervisor node is to generate the approval decision of prover choice contradicting the choice of witness and the location authority of other supervisor nodes. In this scenario, when a decision block will be propagated in the admin layer, then honest supervisor nodes will validate it and reject it, and later, the location proof generated against this decision block will also be rejected. Hence, MobChain remains secure until 51% of the supervisor nodes get compromised by the prover.

The large value of uptime indicates that the witness and the location authority are more trustable and reliable. Uptime will help in mitigating the puppet witness attacks and *Sybil* attacks. We can also specify the minimum threshold for uptime of worker nodes to become eligible for participation in proof generation making the puppet witness attack and *Sybil* attacks harder. Once the priority is being calculated for all, witnesses and *LA* with higher priority are chosen and every *SN* informs the *RRSN* about his choice by sending a decision acknowledgment message *'DAM'* i.e.

$$DM_{SNi} = [ID_{SNi}, T_{SNi}, W_C, LA_C, PReq'] \quad (3a)$$

$$DAM_{SNi} = [DM_{SNi}, Sign_{SNi}(DM_{SNi})] \quad (3b)$$

In expression (3a), $ID_{SNi}$ is the unique identifier of $SN_i$, $T_{SNi}$ is the timestamp of $SN_i$ machine, $W_C$ is the witness chosen and $LA_C$ is the location authority chosen by $SN_i$ for assisting the prover in proof generation. While, $Sign_{SN}(DM_{SN})$ serves the non-repudiation of *SN*(s) and later will be used to validate the final decision block constructed by *RRSN*:

$$DB_i = [\{DAM_{SNi}, \ldots, DAM_{SNN}\}, W_C, LA_C, ID_{RRSN}, T_{DBi}] \quad (4a)$$

$$DB'_i = [DB_i, Sign_{RRSN}(DB_i)] \quad (4b)$$

In expression (4a), final decision block *'DB''* incorporates all *DAM* messages received from *SN*(s), the $W_C$ and $LA_C$ are the chosen witness and location authority respectively, over which more than 50% of *SN(s)* have consensus, $ID_{RRSN}$ identifies the creator of the block, and $T_{DBi}$ is the time of the creation of the block. To make the new decision block part of the decision block chain, its unique id '$ID_{DBi}$' is calculated:

$$ID_{DB'i} = H(H(DB'_{i-1}), DB'_i) \quad (5)$$

Now, the *RRSN* will broadcast the new decision block to all *SN*(s), furthermore, it will construct an approval message *'AMsg'* for prover containing the decision block id, chosen witness and location authority:

$$AMsg = [PReq, ID_{DB'i}, W_C, LA_C, T_{AMsg}] \quad (6a)$$

$$AMsg' = [AMsg, Sign_{RRSN}(AMsg)] \quad (6b)$$

In expression (6a), $T_{AMsg}$ is the time of the creation of the approval message and in expression (6b), the approval message constructed in (6a) is signed by *RRSN* so that prover can validate it to ensure that message is from valid *SN*.

#### ***4.3.2 Witness-Oriented Location Proof Generation Phase***

Working of the witness-oriented location proof generation protocol described in Figure 2. is as follows:

1. Prover requests the location authority $LA_C$ (chosen by the admin layer) to assist him in a proof generation by sending an *LPReq* message.

   $$LPReq = [AMsg', T'P] \quad (7)$$

   where *AMsg'* in eq. (6b) is the approval message from the admin layer after distributed consensus, sent by *RRSN* and $T'_P$ is the time of the request to $LA_C$ by the prover.

2. $LA_C$ first validates the *LPReq* against the *AMsg'* received from *RRSN*. Once validated then performs secure localization to ensure that the prover is physically present on the mentioned location. Finally, $LA_C$ generates the location proof *LP* against *LPReq*,

   $$LP = [ID_{LA}, LPReq, T_{LS}] \quad (8)$$

   where $T_{LS}$ is the time of location statement generation.

3. $LA_C$ then creates the location proof assertion request "$AReq_{LP}$" and sends it to witness $W_C$.

   $$AReq_{LP} = [LP, Sign_{LA}(LP)] \quad (9)$$

4. Witness validates the $AReq_{LP}$ against *AMsg'* to ensure that *LA* is asking for the assertion of valid and then performs the secure localization to ensure that location authority and prover are not colluding, and prover is physically present on the reported location. Once verification is successful, witness creates the assertion statement *"AStat"*

   $$AStat = [ID_{DB'i}, ID_P, ID_{LAC}, ID_{WC}, H(AReq_{LP}), T_{AStat}] \quad (10)$$

where $ID_{DB'i}$ is the decision block id, $ID_P$ is the id of the prover, $ID_{LAc}$ is the id of chosen location authority, $ID_{Wc}$ is the id of the chosen witness, $H(AReq_{LP})$ is the hash of assertion request message, and $T_{AStat}$ is the time of assertion statement generation. Finally, the witness generates the assertion response *"AR"* and sends it back to the location authority.

$$AR = [AStat, Sign_{Wc}(AStat)] \quad (11)$$

Witness now generates the final asserted location proof *"ALP"* and sends it back to *LA*. Location authority validates by verifying the signatures of asserted location proof using the public key of witness to ensure it is from the selected witness. Location authority then provides the asserted location proof to the prover.

$$ALP = [AReq_{LP}, AR, T_{ALP}] \quad (12)$$

where $T_{ALP}$ is the time of asserted location proof creation.

5. Prover does not trust the location authority; therefore, it sends the asserted location proof (provided by location authority) back to witness for verification to ensure the provided location proof is endorsed by the witness. Therefore, on receiving *ALP*, prover issues a verification request *VReq* to witness

$$VReq = [ALP, T_{VPReq}] \quad (13)$$

6. Witness responds the prover by *Yes/No* after validation of *VReq* by sending message *"VR"*

$$V = [R, H(ALP), T_V] \quad (14a)$$

where $R = \{Yes, No\}$ and $T_V$ is a time of generation of $V$

$$VR = [V, Sign_{Wc}(V)] \quad (14b)$$

7. After successful verification of asserted location proof, prover sends the acknowledgement $ACK_{ALP}$ to location authority to end the protocol

$$ACK = [ALP, VR, ID_{DB'i}, T_{ACK}] \quad (15a)$$

where $T_{ACK}$ is the time of acknowledgment generation

$$ACK_{ALP} = [ACK, Sign_P(ACK)] \quad (15b)$$

8. After validation of $ACK_{ALP}$, location authority will end the protocol and send $ACK_{ALP}'$ to *RRSN* after signing it.

$$ACK_{ALP}' = [ACK_{ALP}, Sign_{LA}(ACK_{ALP})] \quad (15c)$$

9. *RRSN* after validation of $ACK_{ALP}'$ will made it part of location provenance chain by broadcasting it in admin layer.

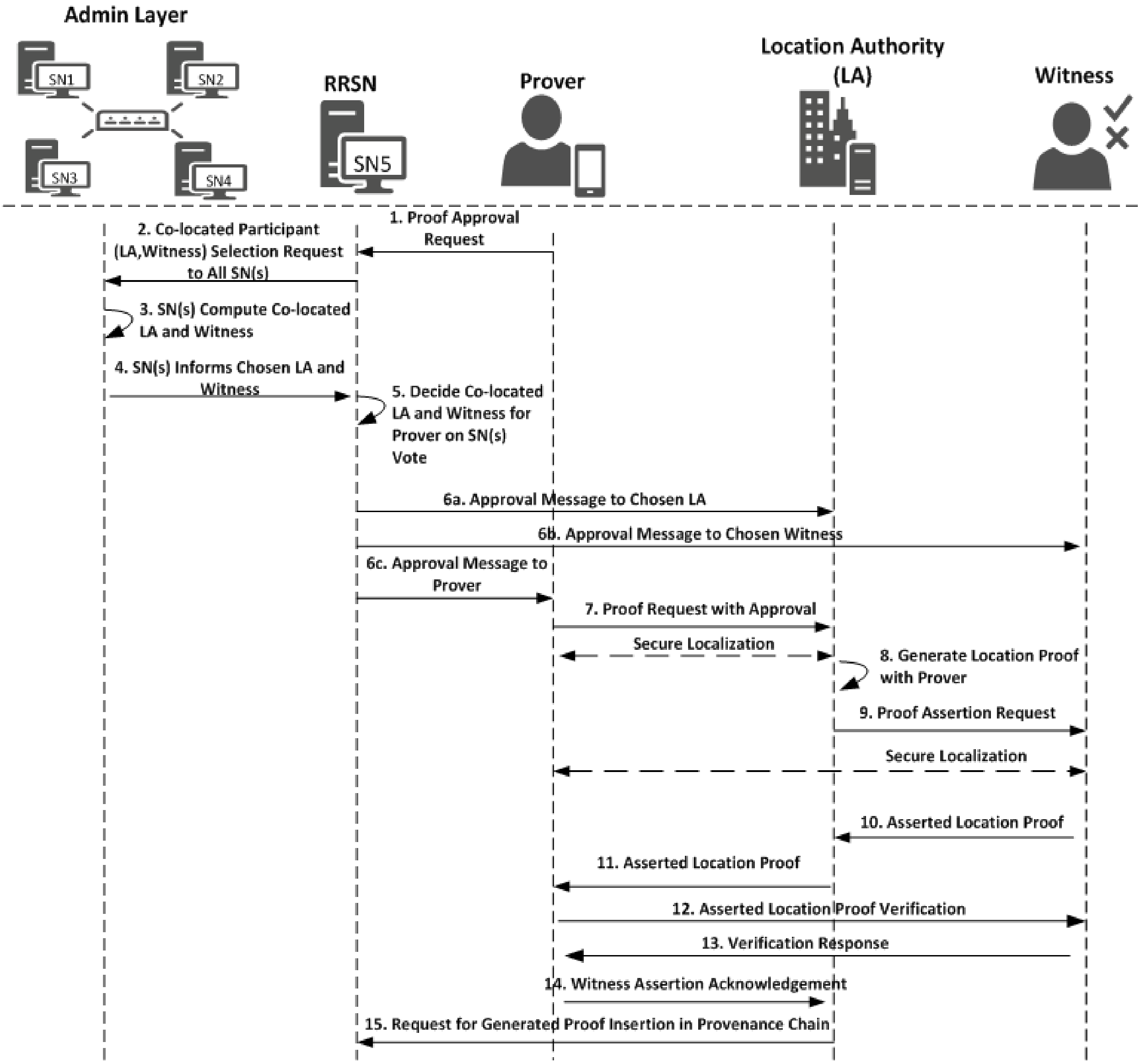

Figure 2. Witness-Oriented Location Proof Protocol

Once the location proof protocol ends, the prover can present the asserted location proof to any third-party service to avail the location-based services. Third-party can check the validity of the generated location proof by requesting any supervisor node of the admin layer.

#### 4.3.3 Colored Petri Net Modeling of MobChain

Colored Petri Nets (CPN) are graphical modeling tools that are suited for modeling communication, synchronization, and concurrency of the systems. The Petri Net modeling is used to: **(a)** simulate and **(b)** provide mathematical representation, and **(c)** analyze the behavior and structural properties of the system [52]. One of the characteristics of CPN modeling is that it is generic, instead of domain specific, which means it is it can model a very broad class of systems characterized as distributed and concurrent. In this section, we will discuss the CPN model of our proposed MobChain architecture. We used CPN Tools version 4.0.1 [53] to transform our MobChain architecture into a CPN model. The CPN model of MobChain is depicted in Figure 3(a) below.

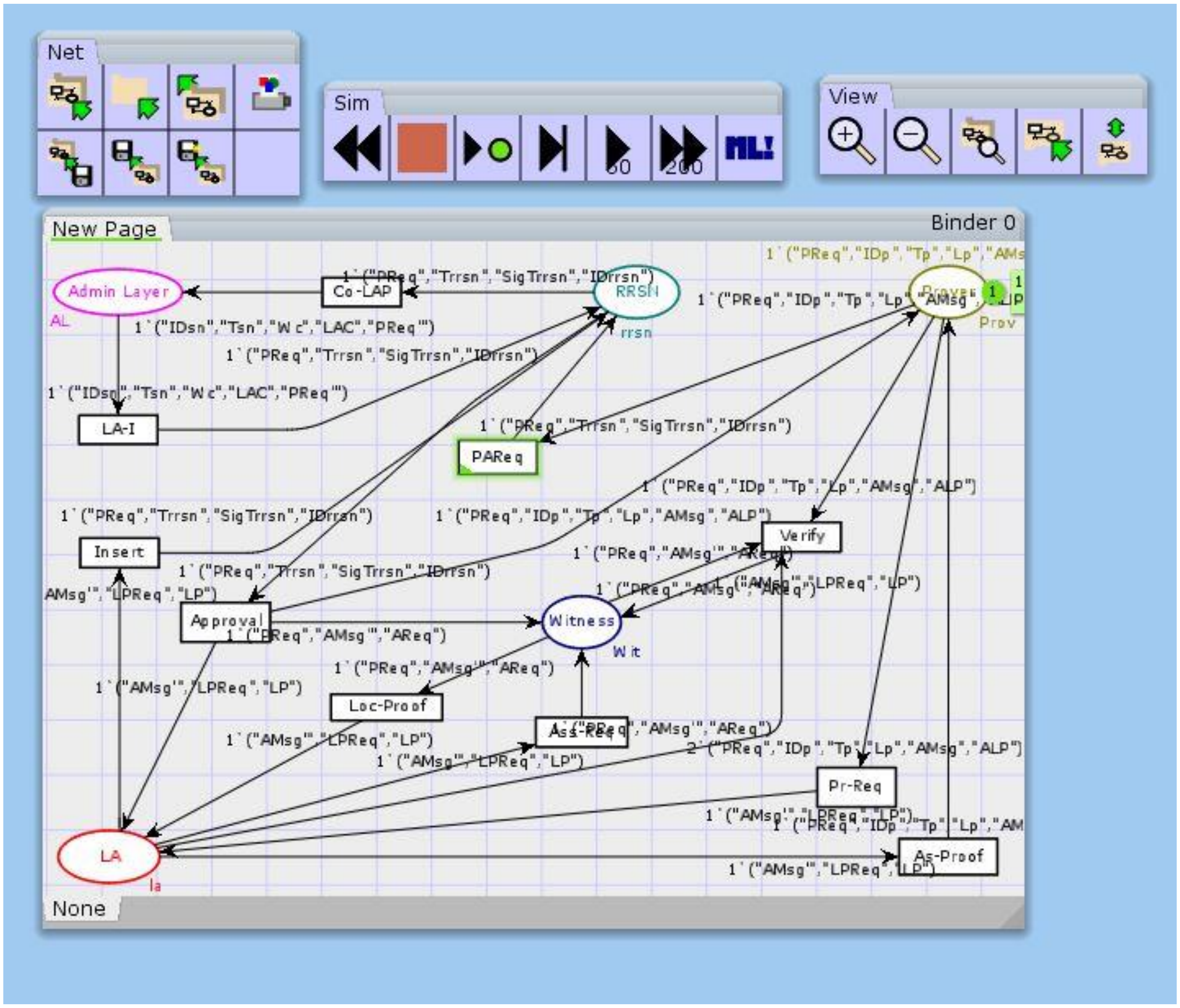

Fig- 3 (a): CPN of the MobChain

A CPN is a Tuple $N = (P, T, A, \Sigma, C, N, E, G, I)$, where P is a set of places, $T$ is a set of transitions, $A$ is a set of arcs, $\Sigma$ is a set of color sets, $C$ is a set of color function that maps P into colors in $\Sigma$, $N$ maps A into $P$ to $T$ or $T$ to $P$, $E$ is an arc expression, $G$ is a guard function, and $I$ is an initialization function. The ovals in the CPN model above represents the places, rectangles represent the transitions, the expressions on the arc represents the tokens (values) consumed from or produced to the places, and the expression on the places represents the initial marking of the place. We started our model with only one (01) token (green circle) in the *Prover* state, as prover is the first state to start the whole process by sending a *PReq* to the supervisor nodes. As seen in the model, initially, only one transition (*PAReq*) is enabled. Once the model starts executing, different transitions are enabled/fired and new markings are observed (as depicted in Fig-3 (b)).

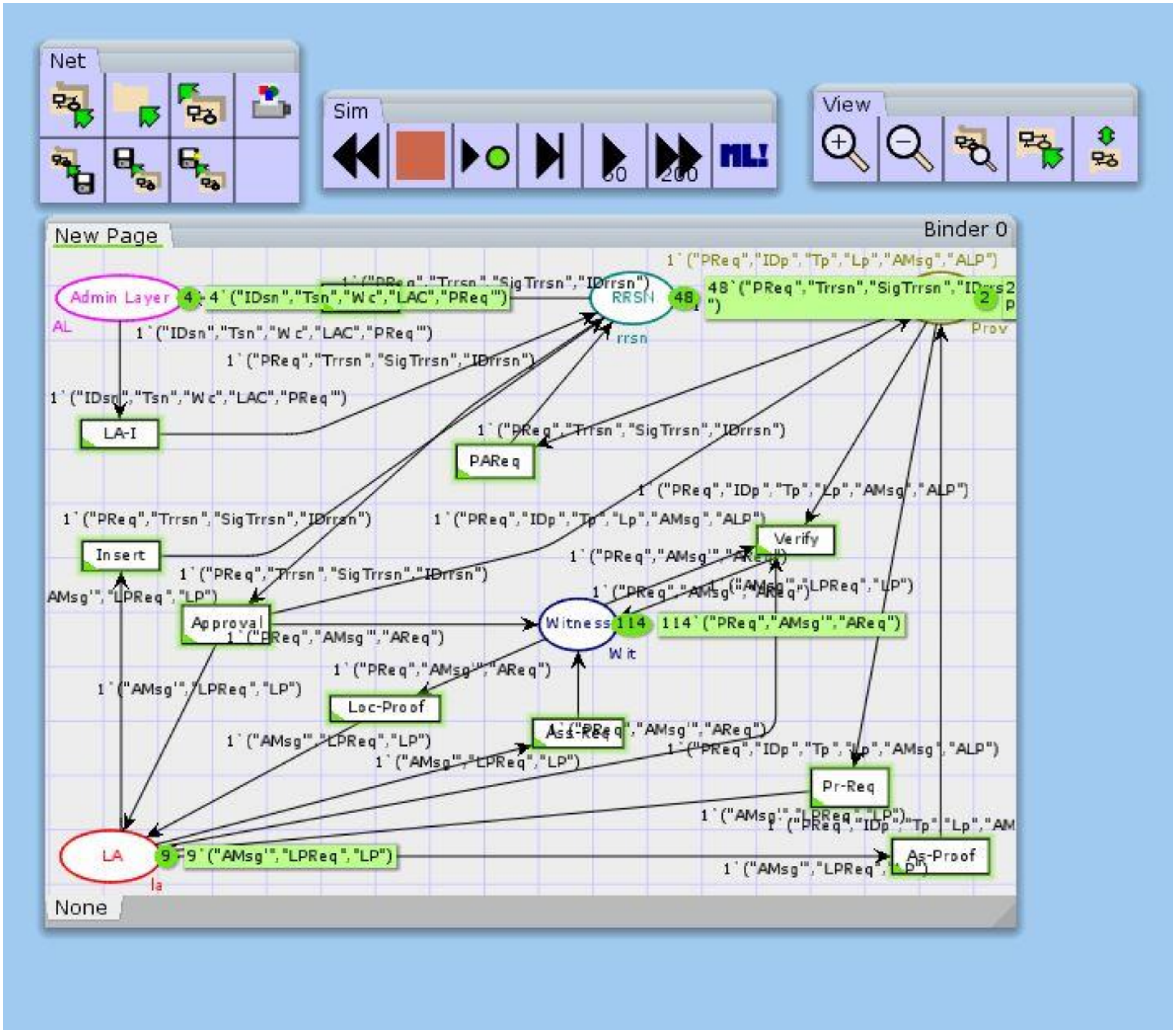

Fig-3 (b): CPN of the MobChain

The CPN model provides new insight about the underlying system, the execution process, and flow of information from a state to another. It verifies the correctness of the and provide evidence if the deadlock exists in a system or not. From the above model, it is evident that all the states of the model are reachable, and every transition of the model is enabled/fired over the course of simulation. Moreover, there is no bottleneck in the aforesaid model, which proves that it is deadlock free. The number of tokens in a state, may vary, depending upon the execution pattern or the number of times a transition is fired. The marking size of the aforesaid model obtained after 1003 steps of simulations are depicted in the Table-2.

Table-2 Marking Size of the states in MobChain

| Marking Size of the States | | | | |
|---|---|---|---|---|
| **State** | **Count** | **Sum** | **Average** | **Std. Deviation** |
| *Admin Layer* | 1003 | 8097 | 8.072782 | 5.318751 |
| *LA* | 1003 | 5355 | 5.338983 | 4.691772 |
| *Prover* | 1003 | 748 | 0.745763 | 1.000085 |
| *RRSN* | 1003 | 20589 | 20.527418 | 16.947553 |
| *Witness* | 1003 | 56707 | 56.537388 | 38.511260 |

## 5. Security Analysis

To highlight the security of MobChain, let see the possibilities for fake proof generation by prover:

1. Suppose all witnesses co-located to the prover are willing for the collusion attack at a time instance. Therefore, the probability of getting malicious witnesses selected is 100%. In this scenario, the fake proof generation is only possible if the chosen location authority is also colluding. If location authority is honest and not willing to collude then the prover will not be able to get the fake proof of location. However, in presence of some honest and malicious location authorities, the probability of a successful collusion attack is still reduced in MobChain as location authority selection control is not in the hands of the prover which is the weakness of past location proof systems.
2. Suppose all available location authorities are willing to collude with the prover. Therefore, whatever location authority is chosen by the admin layer will allow the prover to get fake proof. However, fake proof cannot be generated without having a colluding witness. If a chosen witness is honest then fake proof will not be asserted by the witness and generated proof will be rejected. Since in existing schemes, witness selection control either lies with location authority or prover itself, therefore, three-way collusion cannot be prevented. However, in MobChain, due to the separation of participants select from location proof protocol resists the three-way collusion.
3. All or 51% of the supervisor nodes in the admin layer are compromised and are colluding with prover to assign him the malicious witness and location authority of his own choice for fake location proof generation. We have assumed that the prover is unable to compromise 51% of the supervisor nodes in the admin layer to get the decision of his own choice [25]. Even if the prover is successful to compromise the 51% supervisor nodes, without having colluding location authority and witness, collusion attack is not possible.
4. All witnesses and location authorities in the system available at a time instance are willing to collude with the prover. It means that the whole location proof system is being compromised by the prover. In this situation, even if every supervisor node in the admin layer is honest, whatever witness, and location authority is chosen by the admin layer will allow the prover to generate fake proof successfully following the protocol honestly. However, we assume that at any single instance of time, all the witnesses and location authorities are not willing to collude with the single prover.

It can be deduced from the above discussion that in MobChain collusion attack can only be possible if

1. All available location authorities and witnesses in the system are malicious and willing to collude with the prover any time. Not a single honest location authority and witness available in the system.
2. All or 51% of the supervisor nodes in the admin layer are compromised and the prover takes the decision of his own i.e. colluding location authority and witness are being chosen in the decentralized decision.

Furthermore, to elaborate on the possibilities of attacks, we present a security analysis of MobChain by establishing a matrix indicating the status (honest, malicious) of the participants (Prover, Witness, and *LA*) involved in location proof generation protocol. In Table 3, we are representing the participants of the protocol as *Honest* (H) or *Malicious* (M) and listed the possible attacks in each case.

Table 3. Malicious participants and possible collusion and attacks

| Case # | Prover | LA | Witness | Scenario/ Collusion Class | Threat / Attack | STAMP [8] | WORAL [12] | MobChain |
|---|---|---|---|---|---|---|---|---|
| **1** | H | H | H | Everyone Honest | No attack | ✓ | ✓ | ✓ |
| **2** | H | H | M | Malicious Witness | False endorsement | ✓ | ✓ | ✓ |
| **3** | H | M | H | Malicious LA | Denial of Service, False Assertion | NA | ✓ | ✓ |
| **4** | H | M | M | LA-Witness collusion | Implication Attack | NA | ✓ | ✓ |
| **5** | M | H | H | Malicious Prover, Prover-Prover collusion | False presence, Proof tampering, Sequence alteration, False time, Wormhole/Terrorist Fraud | ✓ | ✓ | ✓ |
| **6** | M | H | M | Prover-Witness collusion | False endorsement | ✓ | ✓ | ✓ |
| **7** | M | M | H | Prover-LA collusion | Puppet Witness Attack | NA | ✓ | ✓ |
| **8** | M | M | M | three-way collusion | Fake proof generation inevitable when all participants are malicious at the same time. | □ | □ | ✓ |

Based on Table 3, we analyze the security of the proposed scheme for each case and explain how it is mitigated.

**Case 1 – All participants are honest:** When all participants are honest, no attack situation exists, and protocol executes normally, and no fake proof generation will be possible.

**Case 2 – Malicious Witness:** Malicious witness may try victimizing the honest prover by endorsing different temporal information than the one in prover request [2] [12]. In MobChain, victimization attacks can be detected on two levels. The first false time endorsement by a witness can be detected by honest location authority and prover by checking $T_{AStat}$ in eq (10) from section 5.3.2 in the assertion response message. *Assertion Response (AR)* in eq (11) is the approval of witnesses about the prover's location $\boldsymbol{L_P}$ (provided in the proof request in eq (1)). Secondly, supervisor nodes will be able to detect the false endorsed proof as prover's request time, and the assertion time $T_{AStat}$ of the endorsing witness is not within a certain range (i.e. few milliseconds). Furthermore, decision block information contains the location proof request time which can also help in validating the spatiotemporal information.

**Case 3 – Malicious Location Authority:** Location authority may try false proof generation for victimizing any prover, but it cannot generate fake proof as no approval is given by the admin layer. To generate the location proof, the location authority needs to include the $AMsg'$ in Eq (6b) to generate the fake proof request. Furthermore, *LA* does not possess the private key for the prover. Besides this, *LA* will not receive a final acknowledgment $ACK_{ALP}$ in eq (15b) from the prover and therefore proof will not be accepted by supervisor nodes. Another possibility is that *LA* may try to deny the proof request for prover deliberately expressing denial of service attack. With MobChain, since multiple location authorities can be deployed on the same site, therefore, the prover can be assigned alternate location authority to aid him in a proof generation.

**Case 4 – Location Authority-Witness Collusion:** Implication attack is an example of location authority and witness collusion [3] [42]. It is a special case of location proof system where innocent prover can be victimized with a false visit claim generated by colluding location authority and witness. Location authority and witness can collude to generate the fake proof to victimize the prover but without approval $AMsg'$ in Eq (6b) from the admin layer and final acknowledgment $ACK_{ALP}$ in eq (15b) from the prover, implication attack is not possible. Reusing or tampering of old $AMsg'$ and $ACK_{ALP}$ can be detected by supervisor nodes because no corresponding decision block will exist in decision blockchain against the tampered $AMsg'$.

**Case 5 – Malicious Prover:** Prover has full control over his smart device, therefore, can override the functionality of the mobile application to tamper the proofs. Attacks possible by malicious prover include *False Presence, False time, Sequence Alteration, Presence Repudiation, Proof Tampering*. In the presence of honest location authority and witness, prover standalone cannot generate fake proof. However, since MobChain maintains the decision blockchain and location provenance, any fake proof generated by malicious prover will be detected and rejected by supervisor nodes. Another possibility is *Prover-Prover Collusion* [8] [50] also known as *Wormhole/Terrorist Fraud* [8] attack can occur when prover A colludes with prover B who is present on the desired physical location to impersonate as A and generate location proof for A. *Terrorist Fraud attack* can easily be detected by honest LA and witnesses due to delay in responses.

**Case 6 – Prover-Witness Collusion:** False endorsement [3] [42] attack is launched by colluding witness and prover. Where prover does not physically co-locate to witness, and witness falsely asserts the location proof to prove the user's physical presence. In the presence of honest LA, false endorsement attacks cannot occur [50]. Secondly, witness selection control is not in the hands of the prover which reduces the probability of such an attack in MobChain.

**Case 7 – Prover-Location Authority Collusion:** Location authority can also be malicious and may collude with the prover. Puppet witness [3] attack is only possible when prover and location authority colludes. The MobChain participant selection mechanism is decentralized, and the prover does not have control over participant's choice, therefore, the probability of a puppet witness attack is zero. Because witness is chosen by the P2P network from registered witnesses and $AMsg'$ in Eq (6b) contains the selected witness. Location proof generated with puppet witness attack can be detected as witness information will not match with witness information in $AMsg'$ as puppet witness is not registered with the admin layer. Such proof will be discarded by supervisor nodes in the admin layer of MobChain.

**Case 8 – Prover-Witness-Location Authority Collusion:** This scenario is known as a *three-way collusion*. To the best of our knowledge, existing schemes have assumed that three-way collusion will not exist. Three-way collusion means the prover, witness, and location authority all 3 can be malicious and maybe colluding at the same time to generate the false proof of location for the prover. The existing schemes have assumed that all 3 parties will not be colluding at the same time, which is a strong assumption [12]. However, in actual if the prover and location authority are colluding then certainly, they can involve the colluding witness. For three-way collusion to be successful, $AMsg'$ in Eq (6b) is required by prover pointing to malicious *LA* and witness of his own choice. Otherwise, without valid $AMsg'$, generate proof will be rejected by supervisor nodes. If prover generates the fake $AMsg'$ then in the final step of location proof generation protocol, it will be detected as no corresponding decision block will exist in the decision blockchain.

## 6. Experimental Evaluation

For experimental evaluation of the scheme, we have developed a proof-of-concept application [55] to simulate the behavior of MobChain. POC is developed using Java AKKA [28] toolkit, an open source library for the development of scalable distributed applications. Simulation results are the average of multiple rounds executed on Hp EliteBook with processor Intel (R) Core (TM) i7-3720QM CPU @ 2.60GHz and 16GB RAM over Windows 10. We used ECC signatures for non-repudiation of messages communicated between entities of the protocol. The following properties of the system are considered to analyze the space requirements and performance of the proposed scheme:

- **Decentralized Decision Time (DDT):** It is the time interval between the location proof request to *RRSN* and the final approval message (created after distributed consensus containing the selected location authority and witness) received by the prover. Measuring decentralized decision time will help to estimate the additional overhead in proof generation time contributed by distributed consensus.
- **Proof Generation Time (PGT):** Proof generation time is the interval between the location proof request generated and the final generated proof received by the prover. It should be short enough (within few seconds) for the system to be practically usable. Proof generation time includes the decentralized decision time.

- **Decision Block Size:** Decision block size depends on the signature scheme used to provide non-repudiation by all entities involved in the decentralized decision. The size of the decision block will have a direct impact on the overall storage capacity required by admin layer nodes as the decision blockchain will be maintained by supervisor nodes.
- **Location Proof Size:** Since location proofs will be stored on the user's mobile device, therefore, its size must be appropriate concerning the storage capacity of smart devices. On the other hand, location proof size also drives the storage capacity of supervisor nodes as the location provenance chain is maintained by the admin layer.

However, the performance of the system and space requirements are directly affected by the following parameters:

1. Number of active workers (Active Witnesses and *LA*(s))
2. Number of supervisor nodes (in a P2P network of admin layer)
    a. Consensus Threshold
3. A key size of a signature scheme

**6.1 Impact analysis of no. of active workers:** In existing schemes such as [3] [12], all co-located witnesses are registered to location authority and no additional computation is required to choose the prover's co-located witness. However, in MobChain, all location authorities and witnesses are registered to supervisor nodes of the underlying P2P network. Therefore, on proof request of the prover, a distributed consensus protocol performs the additional computation to find out the appropriate co-located location authority and witness for prover. Therefore, the number of active workers (location authorities and witnesses) has an impact on distributed consensus time and on overall proof generation time. We have also evaluated the impact of the number of active workers on decision block size and proof size.

**6.1.1 Impact on decentralized decision time and proof generation time:** To see the impact of the number of active workers on decentralized decision time and proof generation time we have performed tests by keeping 15 supervisor nodes in the admin layer of MobChain and consensus threshold set to 8. Furthermore, we used the ECC key size – 224 bits for signed messages communication.

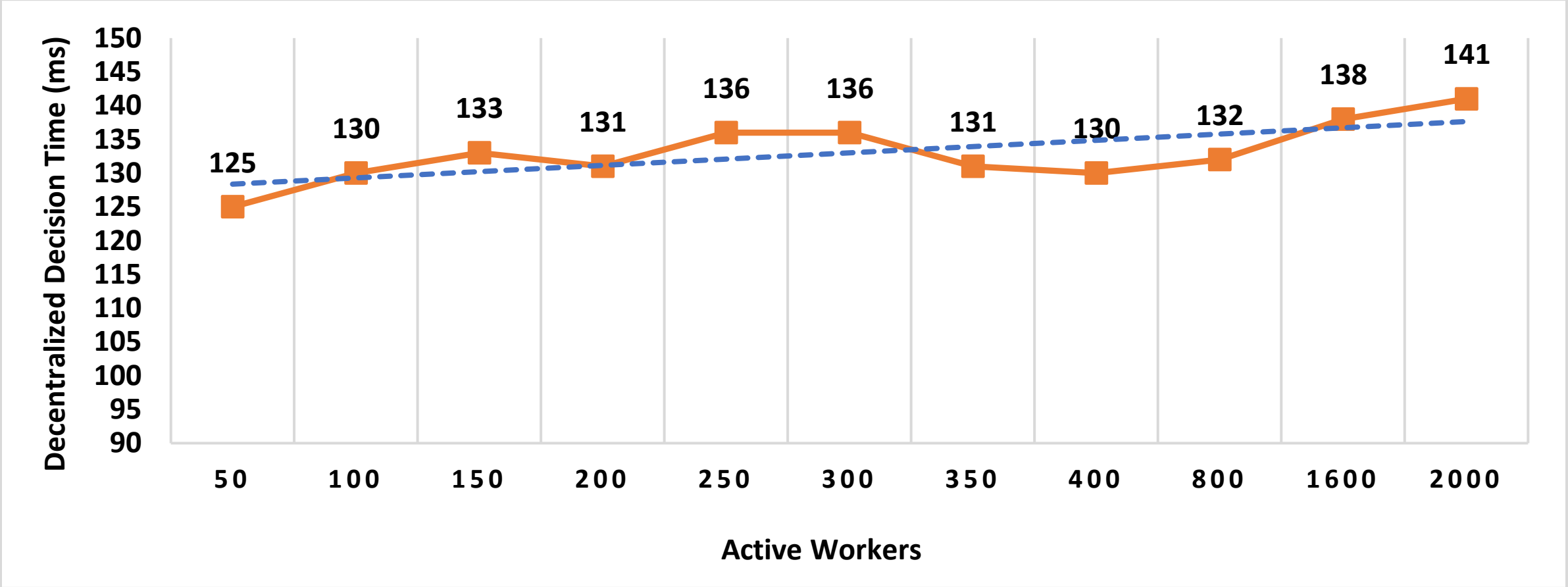

Figure. 4 Active workers impact on decentralized decision time

From Figure 4, we can identify the additional overhead of decentralized decision time in the proof generation process. A slight increase in decentralized decision time is observed with an increase in the number of active workers. In situations with higher worker's count, co-located witness, and location authority's evaluation algorithm might need improvement to reduce this impact of active workers on decentralized decision time. Especially impact of co-located worker's selection algorithm will have a higher impact in peak load times where concurrent requests are received by supervisor nodes in a very short interval of milliseconds. Another factor that can impact the overall performance of decentralized decision time with an increase in the number of active workers is, how frequently worker's geolocation is refreshed in the admin layer P2P network. The specialized cache mechanism can be designed to reduce the time of the co-located worker's evaluation. However, the cache refreshing mechanism demands control over the frequency of worker's geolocation refreshing. The higher the frequency of workers geolocation refreshing, the cache mechanism will start introducing overhead instead of improvement in decentralized decision time. Furthermore, another solution can be the specialized data structure to mitigate the higher frequency of worker's geolocation refresh interval while minimizing the co-located workers evaluation time. In Figure 4, we have plotted the overall proof generation time against the number of active workers to see what are other factors besides decentralized decision time which affect the performance of the scheme.

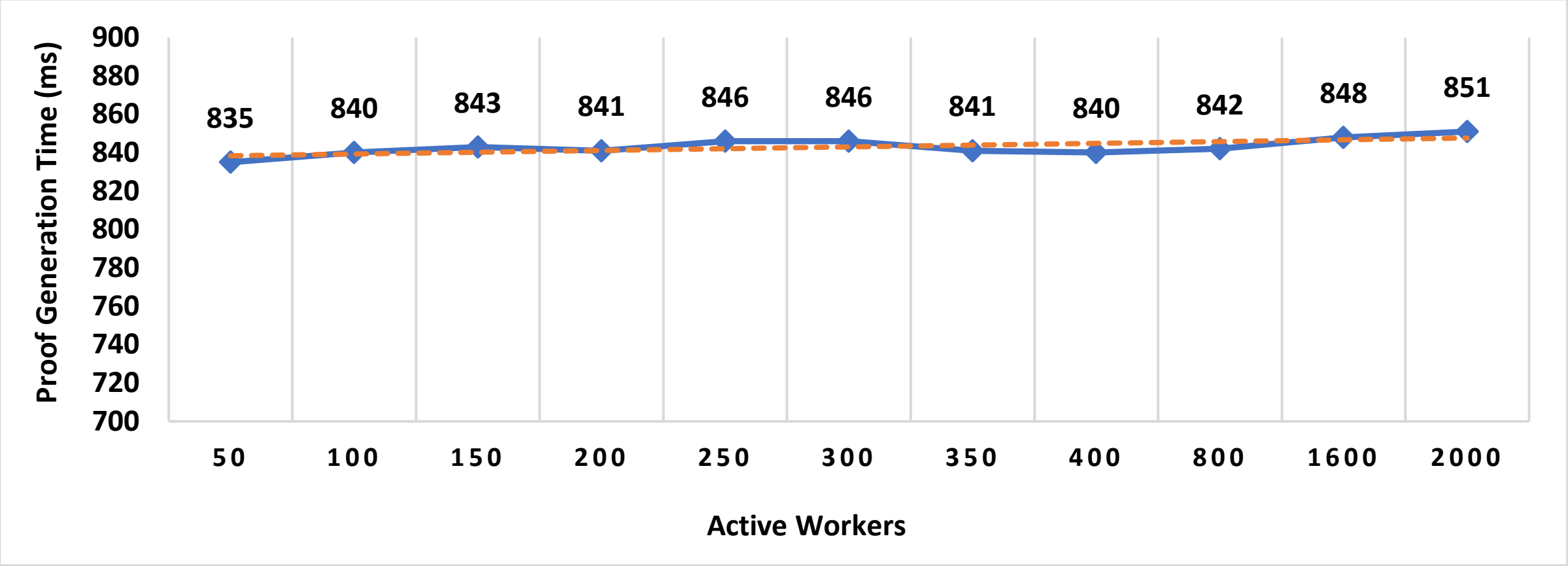

Figure. 5 Active workers impact on proof generation time

By comparing the graph line in Figures 4 and 45, we can deduce that proof generation time has very little overhead of decentralized decision. With experimentation, we have identified that much of the proof generation time includes the secure localization time. Secure localization is done twice in proof generation protocol i) by location authority to identify that prover is physically present in the vicinity ii) by a witness to ensure prover is physically co-located. The distance of participants from the Wi-Fi access point has also little impact on secure localization and message communication. In a real scenario, internet bandwidth and factors affecting the communication will affect the decentralized decision time. However, with our POC simulation, we can identify the lower bounds on the performance.

### 6.1.2 Impact on proof size and decision block size

We measured the size of the location proof generated and the decision block generated by RRSN in the location proof generation process. By flooring the value of size in KB we plotted the graph in Figure 6. Based on the results, we can deduce that the number of active workers does not affect the size of location proof and decision block. Therefore, storage size is independent of the service layer of MobChain for a number of active workers in the system.

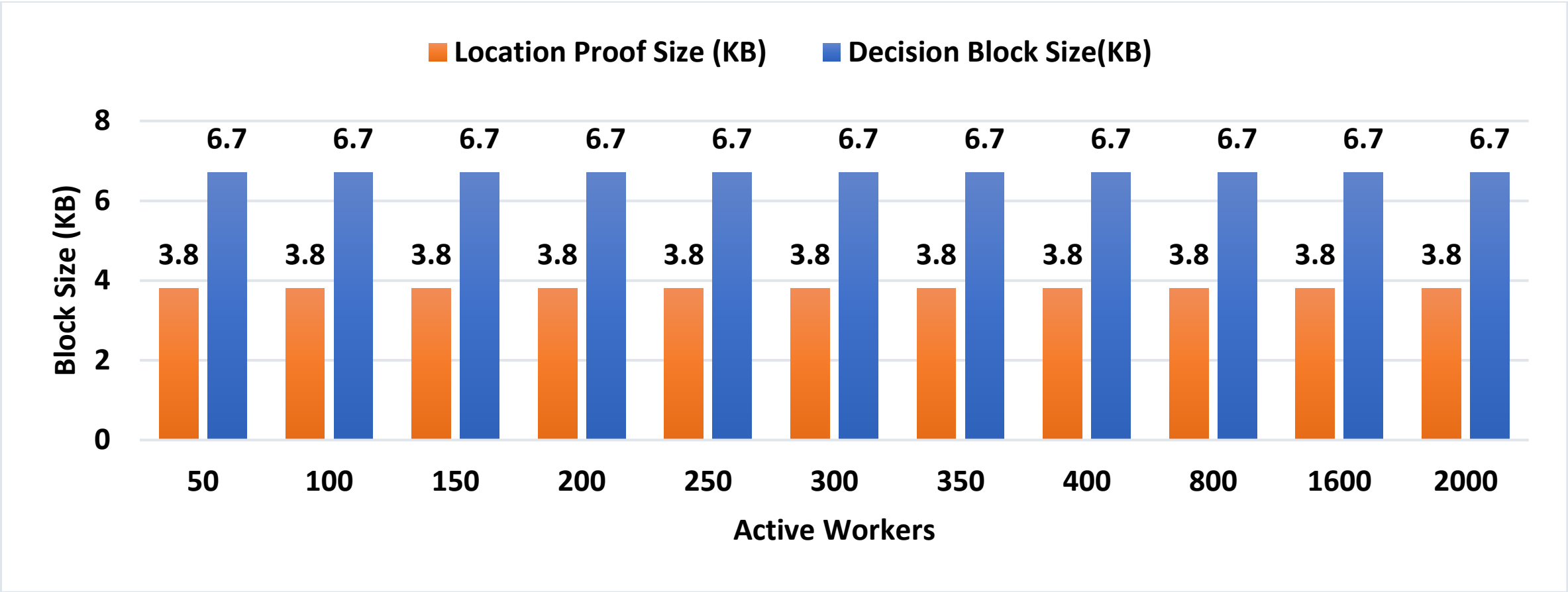

Figure. 6 Active workers impact on proof size and decision block size

## 6.2 No of supervisor nodes impact analysis

In MobChain, the decentralized decision for the selection of participants is introduced as an additional step in location proof protocol. The decentralized decision is taken by supervisor nodes in the admin layer and therefore the size of this P2P network has an impact on decision time and decision block size. However, practically the consensus threshold is the key factor, which controls the impact of many supervisor nodes and determines the reliability of the decision. To measure the impact of the consensus threshold, we performed the experimentation keeping the number of supervisor nodes constant i.e.15, some active workers i.e. 400 and ECC key size – 224 bits.

### 6.2.1 Impact on decentralized decision time and proof generation time

In Figure 7, we have measured the impact of the consensus threshold increase on decentralized decision time and overall proof generation time. As described in the distributed consensus explanation, for consensus threshold value $(N/2) + K$, we gradually increased the value of $K$ and plotted the graph of the results. We observed a slight increase in the decentralized decision time with an increase in consensus threshold value.

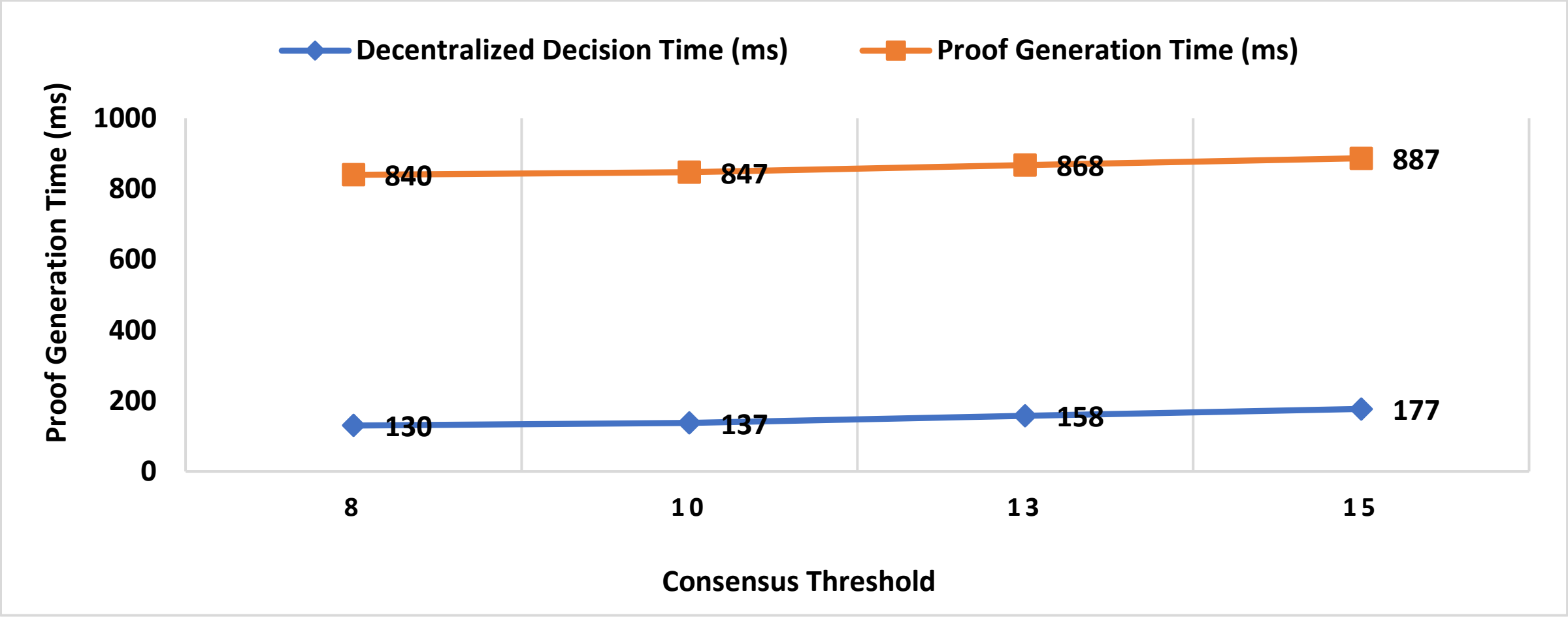

Figure. 6 Consensus threshold impact on decentralized decision time and proof generation time

It means higher the number of supervisor nodes and higher the value of the consensus threshold can increase the decentralized decision time and overall proof generation time. The communication channel between the supervisor nodes is another factor that will enhance the impact on decentralized decision time with an increase in the size of the admin layer network. Furthermore, this impact is also affected by another factor i.e. ECC key size. We repeated the same experiment by increasing the key size of the ECC signature scheme.

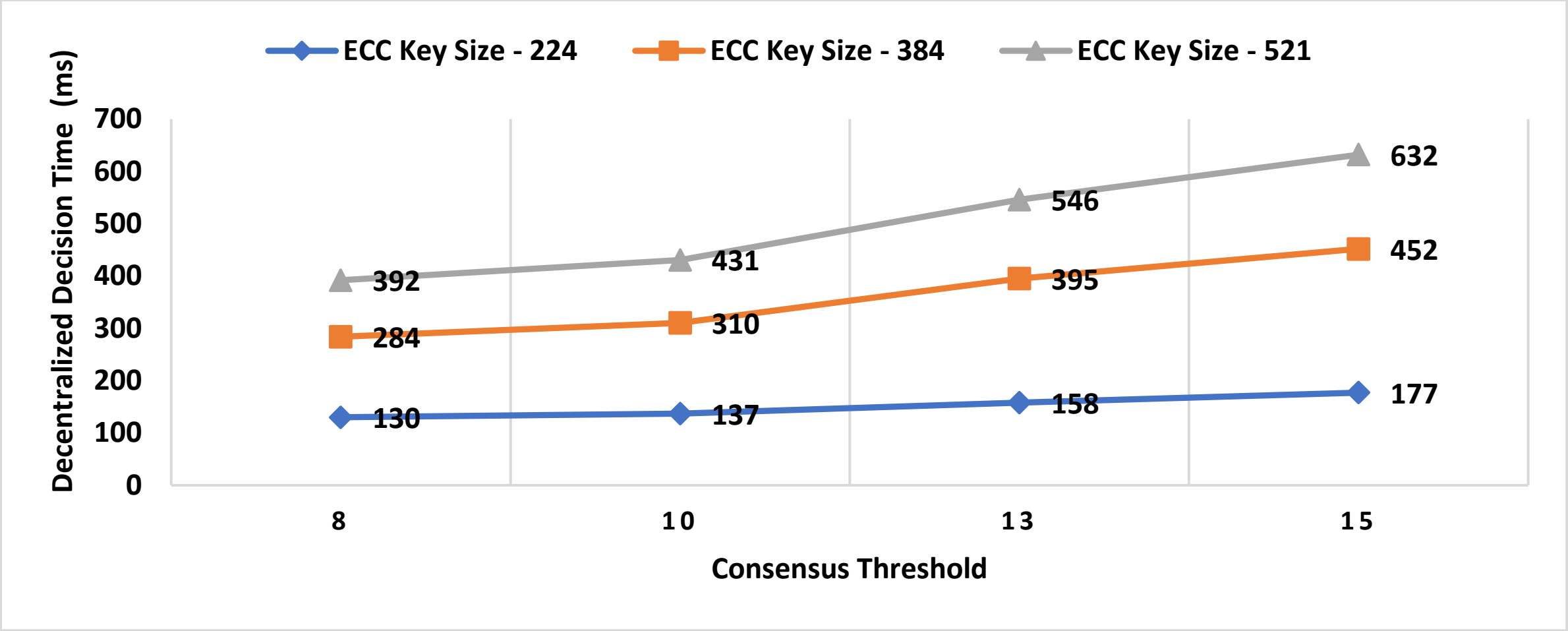

Figure. 8 Consensus threshold and key size impact on decentralized decision time

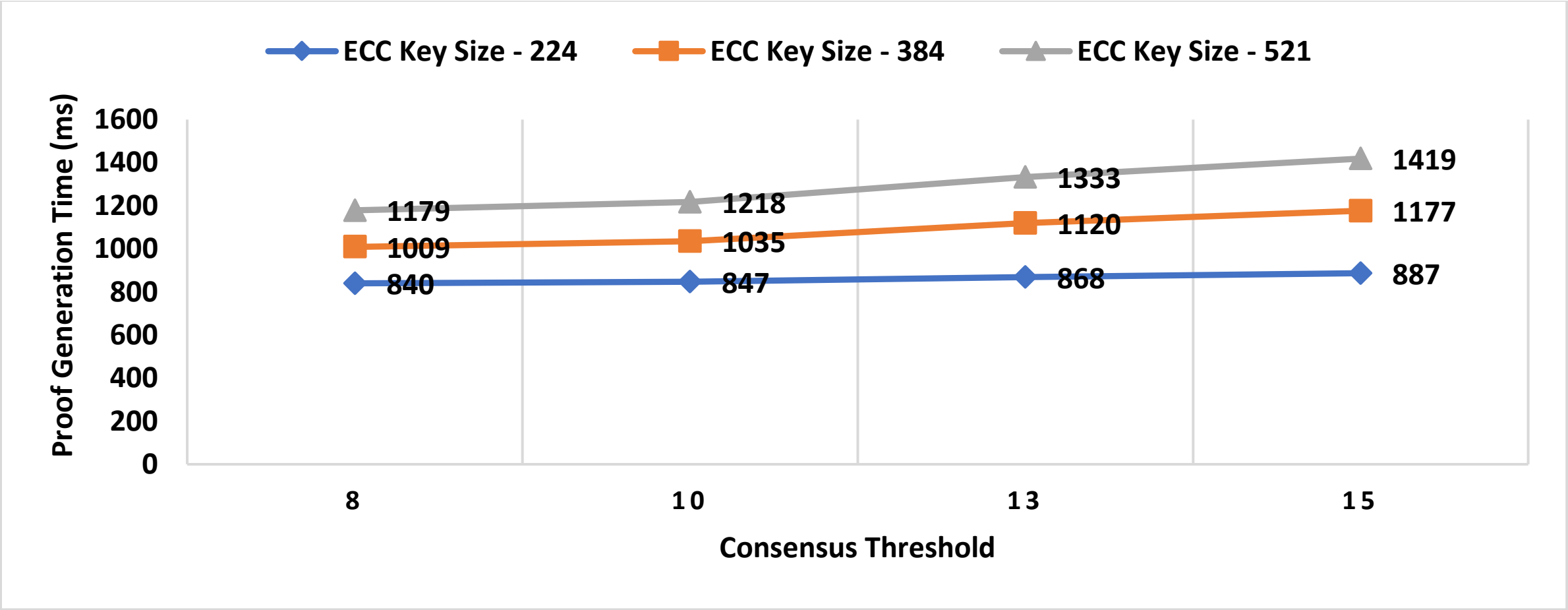

Figure. 9 Consensus threshold and key size impact on proof generation time

From Figures 8 and 9 we can observe the impact of key size on decentralized decision time and overall proof generation time. In comparison with WORAL, ECC key size – 224 bits provides the same level of security as that of RSA with key size 2048 bits, and ECC key size – 521 bits is equivalent to RSA key size – 15360 bits. WORAL proof generation time is ≤ 1 Sec with RSA key size 2048 bits and MobChain with consensus threshold 8 and ECC key size 512 bits is close to it providing better security and three-way collusion resistance. Based on results, we can deduce that the total number of supervisor nodes, the value of the consensus threshold, and the signature scheme's key size collectively determine the decentralized decision time. However, key size also has an impact on the performance of the location proof generation phase after the decentralized decision phase completes. All messages between prover, location authority, and witness are signed during location proof generation protocol.

### 6.2.2 Impact on Location Proof Size & Decision Block Size

Keeping the number of supervisor nodes constant i.e.15, some active workers i.e. 400 and ECC key size – 224 bits, results of consensus threshold impact on location proof size and decision block size are plotted in Figure 10.

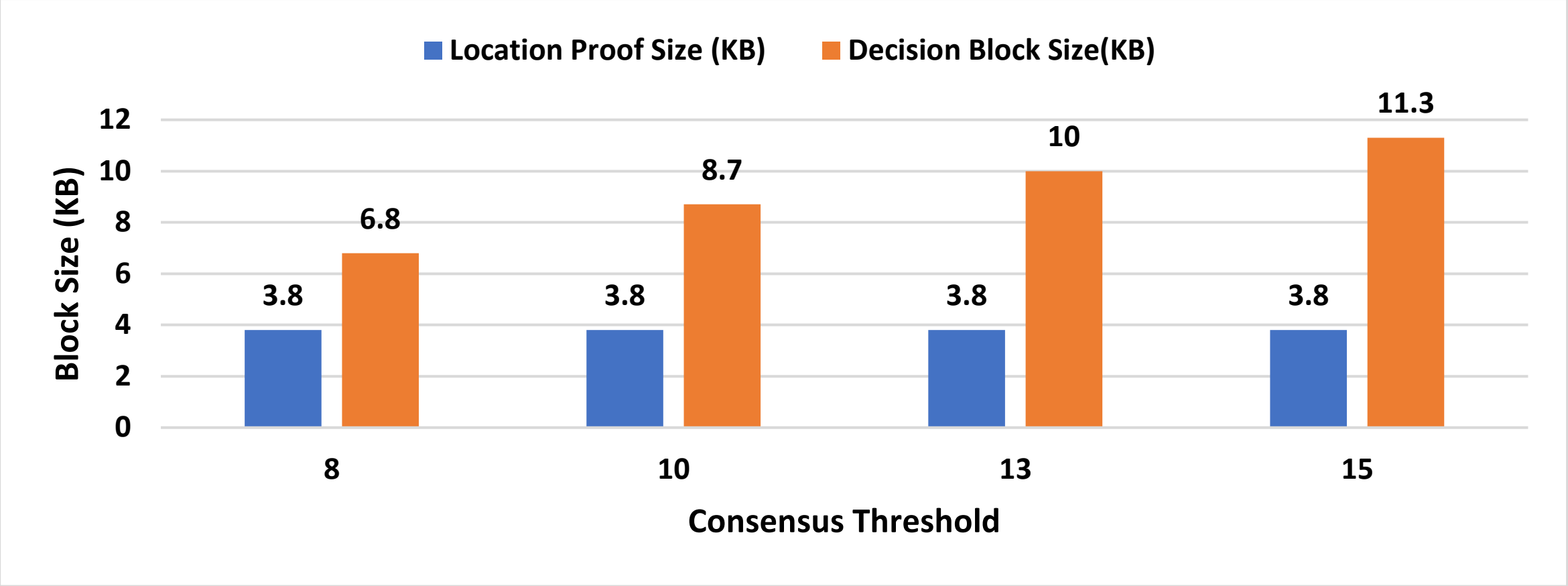

Figure. 10 Consensus threshold impact on location proof size and decision block size

From results in Figure 10, we can deduce that location proof size is independent of the consensus threshold while decision block size is increased with an increase in consensus threshold. Location proof size and decision block are independent of each other. Location proof only contains the decision block ID which is of few bytes. However, the reason for the increase in decision block size is the inclusion of many confirmations of supervisor nodes with information of chosen workers. As soon as 51% of the supervisor nodes develop consensus on the same set of workers for aiding the prover in a proof generation, confirmations received by RRSN up to this time are made part of the decision block. Therefore, in the presence of malicious or malfunction or outdated supervisor nodes, reaching a consensus can vary and it may cause an increase in the size of the decision block. With the help of these confirmations, a decision block is validated by all supervisor nodes before making it part of the decision blockchain on the completion of the protocol. Secondly, the decision block can be validated later to detect any tampering by malicious supervisor node or any corruption. Furthermore, we evaluated the impact of key size along with a consensus threshold on the location proof size and decision block size. Figure 11 shows the results of the location proof size against each key size and consensus threshold value.

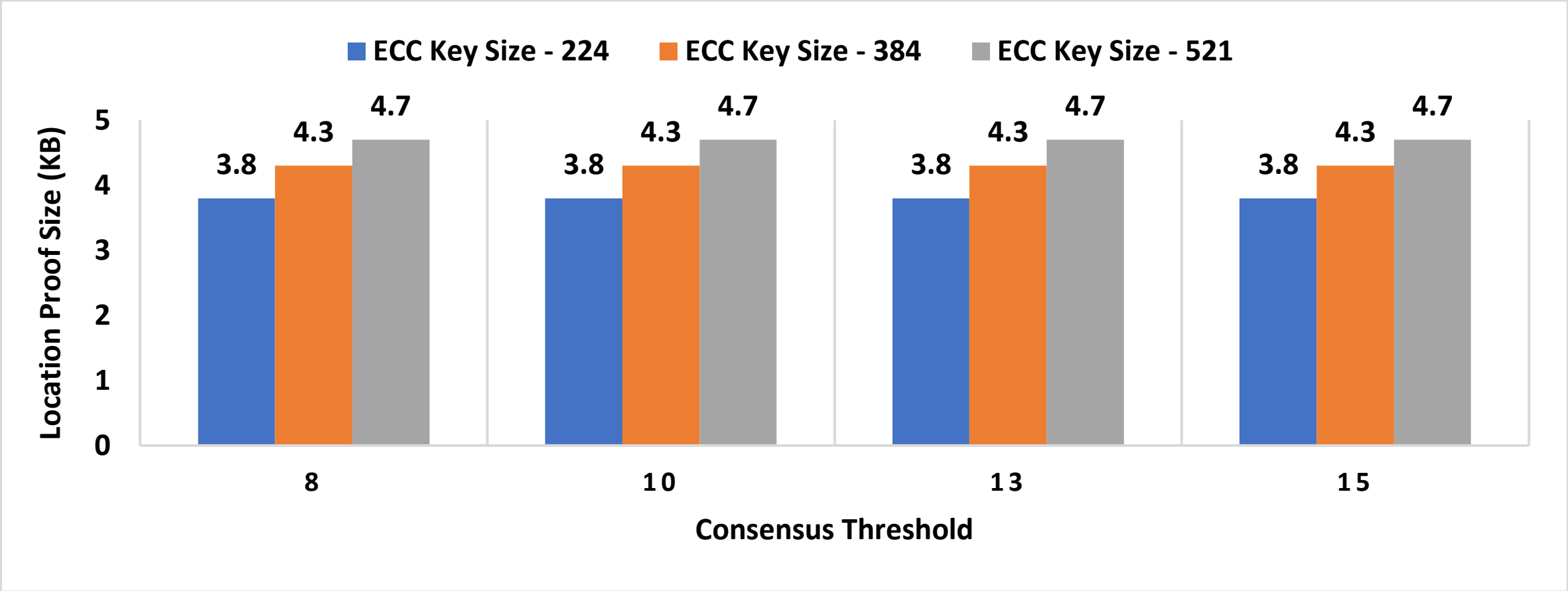

Figure. 11 Consensus threshold and key size impact on location proof size

From the results, it's obvious that location proof size is independent of the consensus threshold however key size increase directly increased the proof size. The reason is the signed information communicated between participants of the protocol during location proof generation. We can deduce the upper and lower bound on the size of proof from Figure 11 results and estimate the storage requirements for smart devices. However, the key size and consensus threshold impact will be higher on the decision block size. We can compare the results of Figure 10 and Figure 12 to see the difference of consensus threshold standalone impact on decision block size and with an increase in key size respectively. An almost 1 KB increase in size is introduced in decision block with an increase in key size. The reason is that with an increase in key size, the signature size included in each confirmation message increases, therefore, increases the overall size of the decision block. Furthermore, the higher the consensus threshold value, the greater number of signed confirmations are included in the decision block resulting from an increase in size.

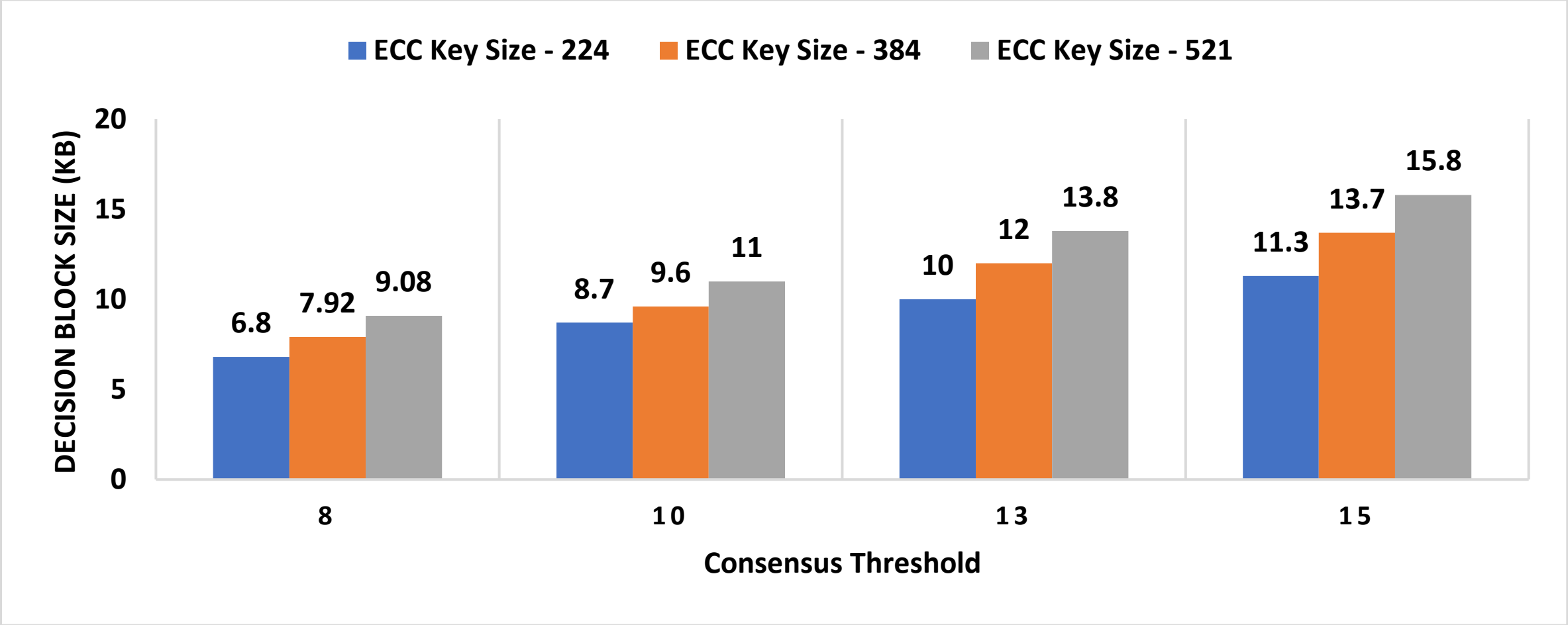

Figure. 12 Consensus threshold and key size impact on decision block size

## 6.3 Concurrent Request Impact on Decentralized Decision Time

Another important factor determining the performance of the MobChain is the number of concurrent requests to supervisor nodes by provers. To see the impact of concurrent requests on performance when all requests are received by the same supervisor node, we performed the tests by keeping the number of supervisor nodes in the admin layer constant i.e. 15 and number of active workers is 1600 and the consensus threshold is 8 with ECC key size 224. The results of the experiment are plotted in Figure 13.

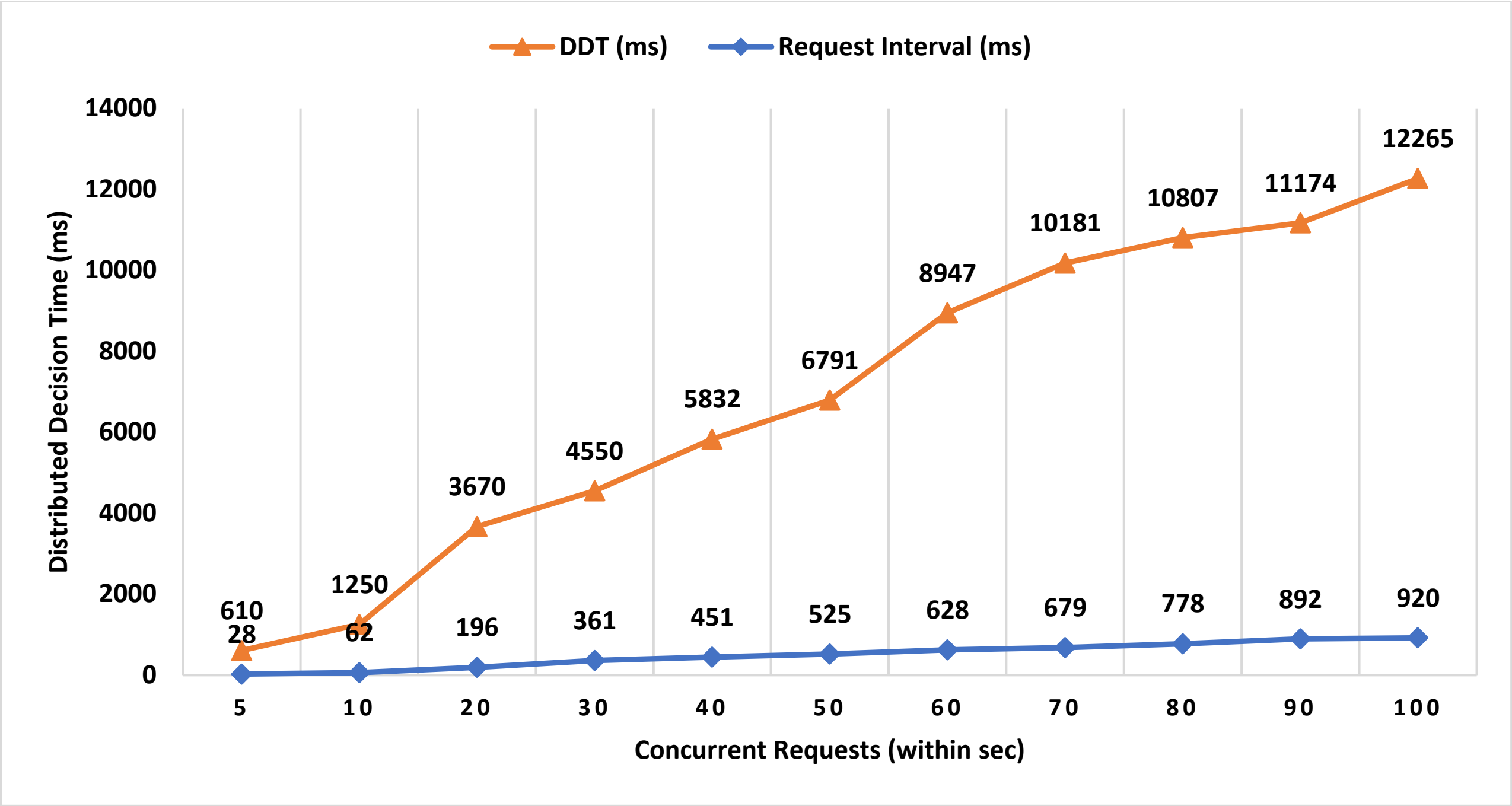

Figure. 13 Concurrent request impact on decentralized decision time

All concurrent requests are generated in less than 1 second on the same supervisor node, to see the impact on decentralized decision time. In Figure 13, *"Request Interval"* is the duration within which all requests are received by RRSN. For example, 5 concurrent requests are received by RRSN within 28 milliseconds and 100 concurrent requests are received within 920 milliseconds. Against each request, RRSN performances the following tasks i) broadcast the location proof request to all admin layer to initiate the distributed consensus against each request ii) compute the co-located workers for prover iii) validate the confirmations received against each proof and wait until the consensus threshold achieved against each request iv) generate decision block once consensus established on the same workers for prover against individual request v) broadcast the decision block to admin layer to make it part of decision blockchain vi) send the approval message including the decision block id against the individual proof request to prover, selected location authority, and witness. Therefore, Figure 13 can give an insight into the impact of concurrent requests load on a single supervisor node and we can see the degradation of performance from results. Hence, we need to re-evaluate the protocol to identify the improvements for reducing the impact of concurrent requests within a short interval on a single supervisor node.

From the results of all the experiments, we can summarize that keeping the ECC key size 224, consensus threshold $(N/2) + 1$, and number of supervisor nodes equal to 15, and assuming the lower number of concurrent requests for the same supervisor nodes within a second, MobChain performance is close to WORAL while providing the resistance to three-way collusion. However, WORAL and MobChain comparison are provided in table 4, showing the cost introduced by MobChain by extending the WORAL to resist three-way collusion.

Table 4. Comparison with existing techniques

| **Feature** | **STAMP [8]** | **WORAL [12]** | **MobChain** |
|---|---|---|---|
| **Proof Generation Time (sec)** | $\leq 3$ | $\leq 1$ | $\leq 1$ |
| **Proof Size (KB)** | $\leq 1.3$ | $\leq 2$ | $\geq 4.7$ |
| **Number of entities involved in the protocol** | Multiple | 3 | 3 |
| **Malicious LA** | Partial | Yes | Yes |
| **Vulnerability (%)** | ≥ 75 | 12.5 | 0 |
| **Decentralized Decision Time (ms)** | NA | NA | ≤ 200 |
| **Decision Block Size (KB)** | NA | NA | ≥ 6.8 |

With optimal parameter configurations, MobChain can compete for WORAL in performance however storage requirements of MobChain are high concerning both smart devices and commodity machines used as supervisor nodes.

## 7. Conclusion and Future work

Witness oriented location proof protocols with location provenance paved the path for improving the real-life business operations. However, the trustworthiness of location proofs is a real concern as participants of the protocol may collude. Trustworthiness cannot be guaranteed without resistance to three-way collusion. MobChain architecture introduced the distributed consensus for location proof generation protocol of witness-oriented schemes to provide resistance against three-way collusion. MobChain proof-of-concept application provides the lower and upper bounds on the performance and storage requirements. Comparisons with the state-of-the-art solutions show that MobChain is computationally efficient, highly available while improving the security of LPS. In the future, we need to transform the proof-of-concept application to proper application to test it in a real environment. Furthermore, we intend to optimize the algorithm of witness and location authority selection to reduce the impact of active workers registered on MobChain network. Furthermore, location proof size and decision block size need to be optimized to reduce the storage requirements, especially on mobile devices. Parallelization techniques along with load balancing mechanisms can also reduce the performance degradation of supervisor nodes under high concurrent requests state.

# Acknowledgment

The authors would like to thank COMSATS University Islamabad, Higher Education Commission (HEC), and Elixir Technologies, Pakistan for their support and encouragement. The authors would also like to thank the anonymous reviewers for their valuable suggestions and insights to improve the quality of our work.